\documentclass[aps]{revtex4}%
\usepackage{amsfonts}
\usepackage{amsmath}
\usepackage{amssymb}
\usepackage{graphicx}

\newcommand{\be}{\begin{equation}}
\newcommand{\ee}{\end{equation}}
\newcommand{\bea}{\begin{eqnarray}}
\newcommand{\eea}{\end{eqnarray}}

\newcommand{\sbspcl}{}
\newcommand{\sbspc}{}
\newcommand{\sspc}{\!\!\!\!\!}

\begin{document}

\title{Phase coexistence in the hard-sphere Yukawa chain fluid with chain length polydispersity:\\
High temperature approximation}

\author{S. P. Hlushak, Yu. V. Kalyuzhnyi}

\address{Institute for Condensed Matter Physics, Svientsitskoho 1, 79011
  Lviv, Ukraine}

\date{\today}

\begin{abstract}

High temperature approximation (HTA) is used to describe the phase behavior of polydisperse
multi-Yukawa hard-sphere chain fluid mixtures with chain length polydispersity.
It is demonstrated that in the frames of the HTA
the model belongs to the class of ``truncatable free energy models'', i.e. the  models
with thermodynamical properties (Helmholtz free energy, chemical potential and pressure)
defined by the finite number of generalized moments. Using this property we were able to
calculate the complete phase diagram (i.e., cloud and shadow curves as well as binodals) and
chain length distribution functions of the coexisting phases.

\end{abstract}

\pacs{~}
\maketitle

\section{Introduction}

In this paper we continue our study of the phase behavior of
polydisperse fluid mixtures. In the previous studies
\cite{my1,my2,my3,my4,my5} a scheme based on integral equation theory
that allows one to include correlation between the particles on the level
of the mean spherical approximation (MSA) has been developed. Usually the
theoretical concepts used to study polydisperse systems treat such systems
as a mixture with an infinite number of components, each of them
characterized by a (continuous) variable $x$, which is distributed according to
the distribution function $f(x)$. In generalizing the formalism of the
MSA for a mixture with a finite number of components to the polydisperse case,
we obtained expressions for the structural and thermodynamic properties that
can be expressed via a finite number of generalized moments of the distribution
function $f(x)$. Such models are called truncatable free energy (TFE) models and
have the attractive feature that the formally infinitely many coexistance
equations for phase equilibrium can be mapped onto a finite number of coupled,
highly non-linear equations in these moments. Polydisperse mixture of hard-sphere
Yukawa \cite{my1,my5} and charged hard-sphere fluid \cite{my2,my3,my4} have been
studied via this approach. In the case of Yukawa mixtures application of the MSA
is restricted to the systems with factorized version of the one-Yukawa interaction,
i.e. the matrix of the coefficients describing the strength of the corresponding interaction
is factorized into the product of two vectors. To remove this restriction recently
high temperature approximation (HTA) for polydisperse multi-Yukawa hard-sphere fluid
has been developed \cite{Kalyuzhnyi2006}. It was demonstrated that within HTA
polydisperse multi-Yukawa hard-sphere model belongs to the family of truncatable
free energy models. Unlike MSA, HTA includes correlations at the
level of the reference system and in general
it is less accurate than MSA, especially at lower densities \cite{henderson,hansen}.
However, HTA can be used to describe the phase behavior of a polydisperse hard-sphere Yukawa
mixture with arbitrary number of non-factorized Yukawa tails. This feature makes the number of the
systems amenable to HTA description much larger than the number of the systems amenable
to MSA description.

In this paper we are focused on the phase behavior of polydisperse mixture of Yukawa hard-sphere
chain molecules with polydispersity in the chain length. Our study is based on the extension
of the recently proposed HTA for the model at hand.
Laplace transforms of the site-site radial distribution functions (RDF) of the reference system,
which are needed as an input, are calculated using product reactant Ornstein-Zernike approach
(PROZA) \cite{proza1,proza2,proza3}, appropriately reformulated in terms of the average
total and direct correlation functions. For thermodynamical properties of the reference system
we have used thermodynamic perturbation theory of Wertheim \cite{w_jcp,Alejandro1997}, generalized
for polydisperse hard-sphere chain mixture.
Similar as in the case of polydisperse hard-sphere Yukawa mixture polydisperse hard-sphere
Yukawa chain mixture within present version of the HTA belongs to the family of TFE models.
This feature allows us to
use a scheme developed earlier \cite{my1,my2,my3,my4} and to calculate
complete phase diagram of the model (i.e., cloud and shadow curves as well as binodals) and
chain-length distribution functions of the coexisting phases.

\section{Hard-sphere Yukawa chain model}

We consider mixture of freely jointed tangent hard-sphere Yukawa chain
molecules of $N$ different chain species with densities $\rho_{a}$
and chain lengths $m_{a}$. Here and below small Latin letters
$a,b,c$ will denote species of chain molecules that take values
$1..N$,  and small letters $i,j,k,...$ will indicate monomer
species of the chain, which take values $1,...,m_a$.
Monomers in each chain are assumed to be of the same size $R^a$. In addition
to the hard-sphere interaction Yukawa potential is acting between monomers,
which are not connected directly.
The resulting
hard-sphere Yukawa potential reads
\bea
u^{ab}_{Y}(r)=\left\{
                \begin{array} {l@{~~}l}
                \infty,  \qquad\qquad\qquad\qquad\qquad\qquad\qquad\quad 0 \le r \le R^{ab} \\
                -\frac{\epsilon_0r_0}{r}e^{-z^{(0)}\left(r-R^{ab}\right)}, \qquad\qquad\qquad\qquad  R^{ab}
                < r \label{Yukawa}
                \end{array} \right.
\eea
where $\epsilon_0$ represents energy at contact, ${z^{(0)}}$ is
Yukawa screening length and $r_0$ is distance unit. For simplicity
reasons we have employed only one-Yukawa interaction, but our
theory can be easily extended by multi-Yukawa potential, which can
mimic large variety of realistic potentials.

\section{Structure properties of the reference system}

For the sake of simplicity instead of the regular site-site RDF $g_{ij}^{ab}(r)$ we are using here
average site-site RDF $g^{ab}(r)$, defined as follows:

\be
\bar {g}^{ab}(r)={1\over m_am_b}\sum_{ij}g_{ij}^{ab}(r).
\label{H_averaged}
\ee

These RDFs are calculated from the averaged version of the multidensity OZ equation, derived
below.

\subsection{Averaged version of the multidensity OZ equation in the ideal chain approximation.}

Multidensity OZ equation for the multicomponent chain fluid in the
ideal chain approximation is \cite{proza1,proza2,proza3}:

\be
\mathbf{h}^{ab}_{ij}(r)= \mathbf{c}^{ab}_{ij}(r) +
\sum_{c,k}\int{\left[\mathbf{c}^{ac}_{ik}(r')+\mathbf{\Delta}^{ac}_{ik}(r')\right]\rho_c\mathbf{\alpha}
\left[\mathbf{h}^{cb}_{kj}(\left|\textbf{r}-\textbf{r}'\right|)+\mathbf{\Delta}^{cb}_{kj}(\left|\textbf{r}-\textbf{r}'\right|)\right]d\textbf{r}'}\label{eqav1}
\ee
where we have extracted from the correlation functions the terms with delta functions and denoted them by
$\mathbf{\Delta}^{ab}_{ij}(r)$. The matrices used in (\ref{eqav1})
are total correlation function matrix
\be
\textbf{h}^{ab}_{ij}(r) = \left(%
\begin{array}{ccc}
  h^{ab}_{ij,00}(r) & h^{ab}_{ij,0A}(r) & h^{ab}_{ij,0B}(r) \\
  h^{ab}_{ij,A0}(r) & h^{ab}_{ij,AA}(r) & h^{ab}_{ij,AB}(r) \\
  h^{ab}_{ij,B0}(r) & h^{ab}_{ij,BA}(r) & h^{ab}_{ij,BB}(r) \\
\end{array}%
\right),
\ee
direct correlation function matrix

\be
\textbf{c}^{ab}_{ij}(r) = \left(%
\begin{array}{ccc}
  c^{ab}_{ij,00}(r) & c^{ab}_{ij,0A}(r) & c^{ab}_{ij,0B}(r) \\
  c^{ab}_{ij,A0}(r) & c^{ab}_{ij,AA}(r) & c^{ab}_{ij,AB}(r) \\
  c^{ab}_{ij,B0}(r) & c^{ab}_{ij,BA}(r) & c^{ab}_{ij,BB}(r) \\
\end{array}%
\right),
\ee
density matrix

\be
\mathbf{\alpha}=\left(%
\begin{array}{ccc}
  1 & 1 & 1 \\
  1 & 0 & 1 \\
  1 & 1 & 0 \\
\end{array}%
\right),
\ee
\noindent
and the matrix containing delta-functions:

\be
\mathbf{\Delta}^{ab}_{ij}(r) =
\delta_{i,j+1}\left(%
\begin{array}{ccc}
  0 & 0 & 0 \\
  0 & 0 & \Delta^{ab}(r) \\
  0 & 0 & 0 \\
\end{array}%
\right) +
\delta_{i,j-1}\left(%
\begin{array}{ccc}
  0 & 0 & 0 \\
  0 & 0 & 0 \\
  0 & \Delta^{ab}(r) & 0 \\
\end{array}%
\right),
\ee
 where
\be
 \Delta^{ab}(r)=\frac{\delta_{ab}}{4\pi\rho_a\left(R^{ab}\right)^2}\delta(r-R^{ab}),\quad\quad
 R^{ab}=(R^a+R^b)/2.
\ee

Taking the average of both sides of equation (\ref{eqav1}) we have

\be
\sspc\frac{1}{m_a m_b}\sum_{ij} \textbf{h}^{ab}_{ij}(r)=
\frac{1}{m_a m_b}\sum_{ij}\textbf{c}^{ab}_{ij}(r) + \frac{1}{m_a
m_b}\sum_{ij}\sum_{ck}\int{\left[\textbf{c}^{ac}_{ik}(r')+\mathbf{\Delta}^{ac}_{ik}(r')\right]\rho_c\mathbf{\alpha}
\left[\textbf{h}^{cb}_{kj}(\left|\textbf{r}-\textbf{r}'\right|)+\mathbf{\Delta}^{cb}_{kj}(\left|\textbf{r}-\textbf{r}'\right|)\right]d\textbf{r}'}
\label{eqav2}.
\ee

By direct multiplication and summation, it can be shown, that

\be
\sum_{ij}\sum_{ck}\mathbf{\Delta}^{ac}_{ik}(r')\rho_c\mathbf{\alpha}\mathbf{\Delta}^{cb}_{kj}(\left|\textbf{r}-\textbf{r}'\right|)=
(m_a-2)\sum_{c}\mathbf{\bar{\Delta}}^{ac}(r')\rho_c\mathbf{\alpha}
\mathbf{\bar{\Delta}}^{cb}(\left|\textbf{r}-\textbf{r}'\right|)
\ee
where we have introduced new matrix

\be
\mathbf{\bar{\Delta}}^{ab}(r) =
\left(%
\begin{array}{ccc}
  0 & 0 & 0 \\
  0 & 0 & \Delta^{ab}(r) \\
  0 & \Delta^{ab}(r) & 0 \\
\end{array}%
\right).\label{newstickyinteraction}
\ee
Similarly

\be
\sum_{ij}\sum_{ck}\mathbf{\Delta}^{ac}_{ik}(r')\rho_c\mathbf{\alpha}\mathbf{h}^{cb}_{kj}(\left|\textbf{r}-\textbf{r}'\right|)=
\sum_{c}(m_c-1)m_b\mathbf{\bar{\Delta}}^{ac}(r')\rho_c\mathbf{\alpha}
\mathbf{\bar{h}}^{cb}(\left|\textbf{r}-\textbf{r}'\right|) \
\label{averaged1},
\ee

\be
\sum_{ij}\sum_{ck}\mathbf{c}^{ac}_{ik}(r')\rho_c\mathbf{\alpha}\mathbf{\Delta}^{cb}_{kj}(\left|\textbf{r}-\textbf{r}'\right|)=
\sum_{c}m_a(m_c-1)\mathbf{\bar{c}}^{ac}(r')\rho_c\mathbf{\alpha}
\mathbf{\bar{\Delta}}^{cb}(\left|\textbf{r}-\textbf{r}'\right|)\label{averaged2},
\ee
where $\mathbf{\bar{c}}^{cb}(r)$ and $\mathbf{\bar{h}}^{cb}(r)$
are averaged direct and total correlation functions
\bea
\mathbf{\bar{c}}^{ab}(r)&=&\frac{1}{m_a
m_b}\sum_{ij}\mathbf{c}^{ab}_{ik}(r),\\
\mathbf{\bar{h}}^{ab}(r)&=&\frac{1}{m_a
m_b}\sum_{ij}\mathbf{h}^{ab}_{ik}(r).
\eea

It should be noted that $\mathbf{\bar{\Delta}}^{ab}(r)$ has a
Kroneker delta symbol $\delta_{ab}$, thus sums over $c$ in
(\ref{averaged1}) and (\ref{averaged2}) can be removed. They
are kept, because they will be used to assemble the OZ
equation from the corresponding averaged quantities.

\bea
\mathbf{\bar{h}}^{ab}(r)&=&\mathbf{\bar{c}}^{ab}(r) +
\sum_{c}\rho_{c}m_c\int\left[\mathbf{\bar{c}}^{ac}(r')\mathbf{\alpha}\mathbf{\bar{h}}^{cb}(\left|\textbf{r}-\textbf{r}'\right|)+
\frac{m_c-1}{m_c}\mathbf{\bar{c}}^{ac}(r')\mathbf{\alpha}\mathbf{\bar{\Delta}}^{cb}(\left|\textbf{r}-\textbf{r}'\right|)+\right.\nonumber\\
&&\left.\frac{m_c-1}{m_c}\mathbf{\bar{\Delta}}^{ac}_{ik}(r')\mathbf{\alpha}\mathbf{\bar{h}}^{cb}_{kj}(\left|\textbf{r}-\textbf{r}'\right|)+
\frac{m_c-2}{m_c}\mathbf{\bar{\Delta}}^{ac}_{ik}(r')\mathbf{\alpha}\mathbf{\bar{\Delta}}^{cb}_{kj}(\left|\textbf{r}-\textbf{r}'\right|)\right]
d\textbf{r}'\label{eqav3}
\eea

Neglecting the terms with
$1/(m_c)^2$
equation (\ref{eqav3}) can be factorized

\be
\mathbf{\bar{h}}^{ab}(r)= \mathbf{\bar{c}}^{ab}(r) +
\sum_{c}\rho_{c}m_c\int{\left[\mathbf{\bar{c}}^{ac}(r')+\frac{m_c-1}{m_c}\mathbf{\bar{\Delta}}^{ac}(r')\right]\mathbf{\alpha}
\left[\mathbf{\bar{h}}^{cb}(\left|\textbf{r}-\textbf{r}'\right|)+\frac{m_c-1}{m_c}\mathbf{\bar{\Delta}}^{cb}(\left|\textbf{r}-\textbf{r}'\right|)\right]d\textbf{r}'}\label{eqav4},
\ee
and all the terms containing delta-functions can
be included into corresponding correlation functions

\be
\mathbf{h}^{ab}(r)= \mathbf{c}^{ab}(r) +
\sum_{c}\rho_{c}m_c\int{\mathbf{c}^{ac}(r')\mathbf{\alpha}
\mathbf{h}^{cb}(\left|\textbf{r}-\textbf{r}'\right|)d\textbf{r}'}\label{eqav5},
\ee
where

\bea
\mathbf{c}^{ab}(r)&=&\mathbf{\bar{c}}^{ac}(r)+\frac{m_c-1}{m_c}\mathbf{\bar{\Delta}}^{ac}(r),\\
\mathbf{h}^{ab}(r)&=&\mathbf{\bar{h}}^{cb}(r)+\frac{m_c-1}{m_c}\mathbf{\bar{\Delta}}^{cb}(r),
\eea
are new averaged correlation functions which now depends only on
chain species index, and are regarded to be the same for all beads
of the chain. Also, new averaged contribution from sticky
interaction (\ref{newstickyinteraction}) was obtained, and it will
be used now in formulation of PPY closure condition in terms of
partial correlation functions $h^{ab}_{\alpha\beta}(r)$ and
$c^{ab}_{\alpha\beta}(r)$

\bea
h^{ab}_{\alpha\beta}(r)&=&-\delta_{\alpha0}\delta_{\beta0},\quad r<R^{ab},\label{closure1}\\
c^{ab}_{\alpha\beta}(r)&=&\frac{\delta_{ab}(m_a-1)}{4\pi\rho_{a}m_{a}\left(R^{ab}\right)^2}\left(\delta_{\alpha{}A}\delta_{\beta{}B}+\delta_{\alpha{}B}\delta_{\beta{}A}\right)\delta\left(r-R^{ab}\right),\quad
r\ge R^{ab}\label{closure2}.
\eea

As one can see the above closure conditions contain factor
$\frac{m_c-1}{m_c}$ which appears due to averaging of the OZ
equation. It appears that the averaged version of the OZ equation
(\ref{eqav5}) coincide with multidensity OZ equation for the fluid
of associating particles forming a chains with the average length
$m_a$ \cite{chang1,chang2}.


\subsection{Solution of the averaged OZ equation}

Solution of the set of the OZ equation (\ref{eqav5}) with closure conditions
(\ref{closure1}) and (\ref{closure2}) was performed employing
Baxter factorization method \cite{Baxter1970}. Factorizing this set of equations (\ref{eqav5})
we have

\bea
\sbspcl
 &&-rc^{ab}_{\alpha\beta}(r)=
\left[q^{ab}_{\alpha\beta}(r)\right]'
-2\pi\sum_{c}\rho_{c}m_c\sum_{\gamma,\delta}(1-\delta_{\gamma\delta}+\delta_{\gamma0}\delta_{\delta0})\frac{\partial{}}{\partial{r}}\int^{min[R^{ac},R^{cb}-r]}_{S^{ac}}{q^{ca}_{\gamma\alpha}(t)
q^{cb}_{\delta\beta}(r+t)dt},\quad\label{eqfact1} \\
&&\qquad\qquad\qquad\qquad\qquad\qquad\qquad\qquad\qquad\qquad\qquad\qquad\qquad\qquad\qquad\qquad\qquad
 for\quad S^{ab}<r<R^{ab},\nonumber\\
\sbspcl
 &&-rh^{ab}_{\alpha\beta}(r)=
\left[q^{ab}_{\alpha\beta}(r)\right]'
-2\pi\sum_{c}\rho_{c}m_c\sum_{\gamma,\delta}(1-\delta_{\gamma\delta}+\delta_{\gamma0}\delta_{\delta0})\int^{R^{ac}}_{S^{ac}}{q^{ac}_{\alpha\gamma}(t)
\left(r-t\right)h^{cb}_{\delta\beta}(\left|r-t\right|)dt},\quad
r>S^{ab},\label{eqfact2}
\eea
where $S^{ab}=\frac{1}{2}\left(R^{(a)}-R^{(b)}\right)$, and $q$
 is Baxter function
\be
q^{ab}_{\alpha\beta}(r)=\left(\frac{1}{2}a^{(a)}_{\alpha}r^2+b^{(a)}_{\alpha}r\right)\delta_{\beta0}+d^{ab}_{\alpha\beta},
\ee
defined in interval $S^{ab}<r<R^{ab}$, with coefficients
$a^{(a)}_{\alpha}$,$b^{(a)}_{\alpha}$ and $d^{ab}_{\alpha\beta}$
found to be
\bea
a^{(a)}_{0}\sbspc&=&\sbspc\frac{1}{\Delta}+\frac{1}{2}\frac{R^{(a)}
    \xi_2}{\Delta^2},\\
b^{(a)}_{0}\sbspc&=&\sbspc-\frac{1}{4}\frac{\left(R^{(a)}\right)^2
    \xi_2}{\Delta^2},\\
d^{ab}_{00}\sbspc&=&\sbspc-\frac{1}{2}\frac{\left(R^{(a)}\right)^2
    \xi_2}{\Delta}+\frac{1}{4}\frac{
    R^{(a)} R^{ab}S^{ab}\xi_2}{\Delta^2},\\
a^{(a)}_{M}\sbspc&=&\sbspc-\frac{1}{2}\frac{(m_a-1)}{m_a\Delta}\left(\delta_{MA}+\delta_{MB}\right),\\
b^{(a)}_{M}\sbspc&=&\sbspc-\frac{1}{2}a^{(a)}_{M}R^{(a)},\\
d^{ab}_{M0}\sbspc&=&\sbspc\frac{1}{2}a^{(a)}_{M}R^{ab}S^{ab},\\
d^{ab}_{0L}\sbspc&=&\sbspc0,\\
d^{ab}_{ML}\sbspc&=&\sbspc\frac{\delta_{ab}}{4\pi\rho^{a}m_{a}R^{(a)}}\left(\frac{m_a-1}{m_a}\right)\left(\delta_{MA}\delta_{LB}+\delta_{LA}\delta_{MB}\right),
\eea
where $\xi_i=\pi\sum_{c}{\rho_{c}m_{c}\left(R^{(c)}\right)^i}$ and
$\Delta=1-\xi_3/6$.

The corresponding expressions for contact values of radial
distribution function
$g^{ab}_{\alpha\beta}(r)=\delta_{\alpha0}\delta_{\beta0}+h^{ab}_{\alpha\beta}(r)$
are as follows:

\bea
g^{ab}_{00}(R^{ab}+)\sbspc&=&\sbspc\frac{1}{\Delta}+\frac{1}{4}\frac{R^{(a)}R^{(b)}\xi_2}{\Delta^2 R^{ab}},\\
g^{ab}_{M0}(R^{ab}+)\sbspc&=&\sbspc-\frac{1}{4}\frac{R^{(b)}}{\Delta R^{ab}}\left(\frac{m_a-1}{m_a}\right),\\
g^{ab}_{0M}(R^{ab}+)\sbspc&=&\sbspc-\frac{1}{4}\frac{R^{(a)}}{\Delta R^{ab}}\left(\frac{m_b-1}{m_b}\right),\\
g^{ab}_{ML}(R^{ab}+)\sbspc&=&\sbspc\frac{\delta_{ab}}{8\pi\rho_{a}m_{a}\left(R^{(a)}\right)^2R^{ab}}\left(\frac{m_a-1}{m_a}\right)^2\left(1-\delta_{ML}\right)
\eea
Note that $g^{aa}_{ML}(r)$ contains also delta function
$\delta\left(r-R^{(a)}\right)$ which describes neighbor hard
spheres of the chain.


\subsection{Laplace transforms of the site-site radial distribution functions}

Integrating both sides of equation
(\ref{eqfact2}) with $\int_{S^{ab}}^{\infty}e^{-sr}...dr$ we have
\bea
&&\sspc\sspc
\sum_{c}\sum_{\delta}\left(\delta_{ca}\delta_{\delta\alpha}-2\pi\rho_{c}m_c\left[Q^{ac}_{\alpha0}(s)+\left(1-\delta_{\delta{}A}\right)Q^{ac}_{\alpha{}A}(s)+
\left(1-\delta_{\delta{}B}\right)Q^{ac}_{\alpha{}B}(s)\right]\right)G^{cb}_{\delta\beta}(s)=\label{lapeq1}\\
&&\sspc\sspc
\delta_{\alpha{}0}\delta_{\beta{}0}\int_{S^{ab}}^{\infty}e^{-sr}dr-{Q^{ab}_{\alpha\beta}}'(s)-2\pi\sum_{c}\rho_{c}m_c
\sum_{\gamma\delta}\left(1-\delta_{\gamma\delta}+\delta_{\gamma{}0}\delta_{\delta{}0}\right)\delta_{\delta{}0}\delta_{\beta{}0}
\int_{dtS^{ac}}^{R^{ac}}{q^{ac}_{\alpha\gamma}(t)\left(\frac{1}{s^2}-\frac{S^{ab}}{s}-\frac{t}{s}\right)e^{-sS^{ab}}}\nonumber
\eea
where Laplace transforms and their derivatives are denoted by capital letters:

\bea
\sspc\sspc
Q^{ab}_{\alpha\beta}(s)\sbspc&=\sbspc&\int_{S^{ab}}^{\infty}e^{-sr}q^{ab}_{\alpha\beta}(r)dr=
e^{-sS^{ab}}\left(\left[a^{(a)}_{\alpha}\left(\varphi_{2}(R^{(b)})+\frac{R^{(b)}}{2}\varphi_{1}(R^{(b)})\right)+\left(\frac{a^{(a)}_{\alpha}R^{(a)}}{2}+b^{(a)}_{\alpha}\right)\varphi_{1}(R^{(b)})\right]\delta_{\beta0}\right.\\
&&\left.+\left(1-\delta_{\alpha0}\right)\left(1-\delta_{\beta0}\right)d^{ab}_{\alpha\beta}\varphi_{0}(R^{(b)})\right),\nonumber\\
\sspc\sspc{Q^{ab}_{\alpha\beta}}'(s)\sbspc&=\sbspc&\int_{S^{ab}}^{\infty}e^{-sr}{q^{ab}_{\alpha\beta}}'(r)dr=
e^{-sS^{ab}}\left(\left[a^{(a)}_{\alpha}\left(\varphi_{1}(R^{(b)})+\frac{R^{(b)}}{2}\varphi_{0}(R^{(b)})\right)+\left(\frac{a^{(a)}_{\alpha}R^{(a)}}{2}+b^{(a)}_{\alpha}\right)\varphi_{0}(R^{(b)})\right]\delta_{\beta0}\right.\\
&&\left.-\left(1-\delta_{\alpha0}\right)\left(1-\delta_{\beta0}\right)d^{ab}_{\alpha\beta}\right),\nonumber
\eea
where
\bea
\varphi_{2}(R^{(b)})\sbspc&=\sbspc&\frac{1-sR^{(b)}+s^2{R^{(b)}}^2/2-e^{-sR^{(b)}}}{s^3},\\
\varphi_{1}(R^{(b)})\sbspc&=\sbspc&\frac{1-sR^{(b)}-e^{-sR^{(b)}}}{s^2},\\
\varphi_{0}(R^{(b)})\sbspc&=\sbspc&\frac{1-e^{-sR^{(b)}}}{s},
\eea
and
\be
G^{ab}_{\alpha\beta}(s)=\int_{S^{ab}}^{\infty}e^{-sr}g^{ab}_{\alpha\beta}(r)dr.
\ee

Similarly as in
\cite{Blum,Golovko,Protsykevich},
solution of equation (\ref{lapeq1}) reduces to inverting
Jacobi like matrix. Note that here we have infinite number of
vectors forming the matrix. To describe briefly our method of
matrix inversion, let us start with equation (\ref{lapeq1})
rewritten in following  form:

\bea
\sum_{c}\sum_{\delta}M^{ac}_{\alpha\delta}G^{cb}_{\delta\beta}(s)
=\hat{a}^{(a)}_{\alpha}\left[\frac{1}{s^2}+\frac{R^{(b)}}{2s}\right]\delta_{\beta0}+\hat{c}^{(a)}_{\alpha}\frac{1}{s}\delta_{\beta0}+\sum_{t}\hat{e}^{(a)}_{\alpha}\frac{\delta_{tb}\delta_{\beta{}A}}{2\pi\rho_{b}m_b}e^{-sR^{(b)}/2}+\sum_{t}\hat{g}^{(a)}_{\alpha}\frac{\delta_{tb}\delta_{\beta{}B}}{2\pi\rho_{b}m_b}e^{-sR^{(b)}/2},
\label{lapeq2}
\eea
where Jacobi like matrix  $M^{ac}_{\alpha\delta}$ is given by

\be
M^{ac}_{\alpha\delta}=\delta_{ac}\delta_{\alpha\delta}-\hat{a}^{(a)}_{\alpha}\hat{b}^{(c)}_{\delta}-\hat{c}^{(a)}_{\alpha}\hat{d}^{(c)}_{\delta}-\sum_{t}\left(\hat{e}^{(a)(t)}_{\alpha}\hat{f}^{(c)(t)}_{\delta}+\hat{g}^{(a)(t)}_{\alpha}\hat{h}^{(c)(t)}_{\delta}\right).\label{matrix1}
\ee
where

\bea
\sbspc&\sbspc&\hat{a}^{(a)}_{\alpha}=e^{-sR^{(a)}/2}a^{(a)}_{\alpha},\\
\sbspc&\sbspc&\hat{b}^{(c)}_{\delta}=2\pi\rho_{c}m_c\left[\varphi_{2}(R^{(c)})+\frac{R^{(c)}}{2}\varphi_{1}(R^{(c)})\right]e^{sR^{(c)}/2},\\
\sbspc&\sbspc&\hat{c}^{(a)}_{\alpha}=e^{-sR^{(a)}/2}\left[\frac{R^{(a)}}{2}a^{(a)}_{\alpha}+b^{(a)}_{\alpha}\right]=e^{-sR^{(b)}/2}\frac{\delta_{\alpha0}R^{(a)}}{2\Delta},\\
\sbspc&\sbspc&\hat{d}^{(c)}_{\delta}=2\pi\rho_{c}m_c\varphi_{1}(R^{(c)})e^{sR^{(c)}/2},\\
\sbspc&\sbspc&\hat{e}^{(a)(t)}_{\alpha}=e^{-sR^{(a)}/2}\frac{m_a-1}{2m_{a}R^{(a)}}\delta_{at}\delta_{\alpha{}B},\\
\sbspc&\sbspc&\hat{f}^{(c)(t)}_{\delta}=\varphi_{0}(R^{(c)})e^{sR^{(c)}/2}\delta_{ct}\left(1-\delta_{\delta{}A}\right),\\
\sbspc&\sbspc&\hat{g}^{(a)(t)}_{\alpha}=e^{-sR^{(a)}/2}\frac{m_a-1}{2m_{a}R^{(a)}}\delta_{at}\delta_{\alpha{}A},\\
\sbspc&\sbspc&\hat{h}^{(c)(t)}_{\delta}=\varphi_{0}(R^{(c)})e^{sR^{(c)}/2}\delta_{ct}\left(1-\delta_{\delta{}B}\right),
\quad for \quad t=1..N.
\eea
The last four vectors $\hat{e}^{(a)(t)}_{\alpha}$,
$\hat{f}^{(c)(t)}_{\delta}$, $\hat{g}^{(a)(t)}_{\alpha}$ and
$\hat{h}^{(c)(t)}_{\delta}$ appear due to decomposition of the
Kronecker delta
\be
\delta_{ab}=\sum_{t}\delta_{at}\delta_{tb}
\ee
in $d^{ab}_{\alpha\beta}$. This decomposition makes matrix
$M^{ac}_{\alpha\delta}$ Jacobi-like. In general it appears to
be impossible to revert such matrix with arbitrary number of
forming vectors $\hat{e}^{(a)(t)}_{\alpha}$,
$\hat{f}^{(c)(t)}_{\delta}$, $\hat{g}^{(a)(t)}_{\alpha}$ and
$\hat{h}^{(c)(t)}_{\delta}$ for $t=1..N$ without notion that some
of these vectors are orthogonal and other can be composed back to
form $\delta_{ab}$:
\bea
\sbspc&\sbspc&\left(\mathbf{\hat{e}}^{(p)}\mathbf{\hat{f}}^{(q)}\right)=\sum_{c}\sum_{\gamma}\hat{e}^{(c)(p)}_{\gamma}\hat{f}^{(c)(q)}_{\gamma}=\frac{m_p-1}{2m_{p}R^{(p)}}\varphi_{0}(R^{(p)})\delta_{pq},\\
\sbspc&\sbspc&\left(\mathbf{\hat{e}}^{(p)}\mathbf{\hat{h}}^{(q)}\right)=\sum_{c}\sum_{\gamma}\hat{e}^{(c)(p)}_{\gamma}\hat{h}^{(c)(q)}_{\gamma}=0,\\
\sbspc&\sbspc&\left(\mathbf{\hat{g}}^{(p)}\mathbf{\hat{f}}^{(q)}\right)=\sum_{c}\sum_{\gamma}\hat{g}^{(c)(p)}_{\gamma}\hat{f}^{(c)(q)}_{\gamma}=0,\\
\sbspc&\sbspc&\left(\mathbf{\hat{g}}^{(p)}\mathbf{\hat{h}}^{(q)}\right)=\sum_{c}\sum_{\gamma}\hat{g}^{(c)(p)}_{\gamma}\hat{h}^{(c)(q)}_{\gamma}=\frac{m_p-1}{2m_{p}R^{(p)}}\varphi_{0}(R^{(p)})\delta_{pq}.
\eea
The inverse matrix was searched in the following form:
\bea
\sspc\sspc
{M^{xy}_{\gamma\lambda}}^{-1}\sbspc&=\sbspc&\delta_{xy}\delta_{\gamma\lambda}+\alpha_{ab}\hat{a}^{(x)}_{\gamma}\hat{b}^{(y)}_{\lambda}+\alpha_{cd}\hat{c}^{(x)}_{\gamma}\hat{d}^{(y)}_{\lambda}+\alpha_{cb}\hat{c}^{(x)}_{\gamma}\hat{b}^{(y)}_{\lambda}+\alpha_{ad}\hat{a}^{(x)}_{\gamma}\hat{d}^{(y)}_{\lambda}
+\sum_{t}\hat{a}^{(x)}_{\gamma}\left(\alpha^{(t)}_{af}\hat{f}^{(y)(t)}_{\lambda}+\alpha^{(t)}_{ah}\hat{h}^{(y)(t)}_{\lambda}\right)\nonumber\\
\sspc\sspc\sbspc&+\sbspc&\sum_{t}\hat{c}^{(x)}_{\gamma}\left(\alpha^{(t)}_{cf}\hat{f}^{(y)(t)}_{\lambda}+\alpha^{(t)}_{ch}\hat{h}^{(y)(t)}_{\lambda}\right)+\sum_{t}\left(\alpha^{(t)}_{eb}\hat{e}^{(x)(t)}_{\gamma}+\alpha^{(t)}_{gb}\hat{g}^{(x)(t)}_{\gamma}\right)\hat{b}^{(y)}_{\lambda}+\sum_{t}\left(\alpha^{(t)}_{ed}\hat{e}^{(x)(t)}_{\gamma}+\alpha^{(t)}_{gd}\hat{g}^{(x)(t)}_{\gamma}\right)\hat{d}^{(y)}_{\lambda}\nonumber\\
\sspc\sspc\sbspc&+\sbspc&\sum_{t_1}\sum_{t_2}\alpha^{(t_1)(t_2)}_{ef}\hat{e}^{(x)(t_1)}_{\gamma}\hat{f}^{(y)(t_2)}_{\lambda}+\sum_{t_1}\sum_{t_2}\alpha^{(t_1)(t_2)}_{gh}\hat{g}^{(x)(t_1)}_{\gamma}\hat{h}^{(y)(t_2)}_{\lambda}+\sum_{t_1}\sum_{t_2}\alpha^{(t_1)(t_2)}_{eh}\hat{e}^{(x)(t_1)}_{\gamma}\hat{h}^{(y)(t_2)}_{\lambda}\\
\sspc\sspc\sbspc&+\sbspc&\sum_{t_1}\sum_{t_2}\alpha^{(t_1)(t_2)}_{gf}\hat{g}^{(x)(t_1)}_{\gamma}\hat{f}^{(y)(t_2)}_{\lambda}\nonumber
\eea
where $\alpha_{ab}$, $\alpha_{cd}$,...,$\alpha^{(t_1)(t_2)}_{gf}$
are unknown coefficients that follow from the relation
$\mathbf{M}\mathbf{M^{-1}}=\mathbf{I}$ with $\mathbf{I}$ being
unity matrix. We will skip these cumbersome calculations.

After receiving unknown coefficients $\alpha$ the inverse matrix
was multiplied by the right hand side of equation (\ref{lapeq2}) giving the following expressions for
the Laplace transforms of the RDFs:

\bea
\sspc\sspc
G^{ab}_{\alpha\beta}(s)\sbspc&=\sbspc&\frac{e^{-sR^{ab}}}{\Delta{}D}\delta_{\alpha0}\delta_{\beta0}\left[\frac{1}{s^2}+\frac{R^{ab}}{s}-R^{a}R^{b}\sum_{c}\frac{\pi\rho_{c}m_c}{\Delta{}s^2}\frac{K^{(c)}}{1-K^{(c)}}R^{(c)}\frac{\varphi_{1}(R^{(c)})}{\varphi_{0}(R^{(c)})}s+\frac{R^{(a)}R^{(b)}\xi_2}{4\Delta{}s}
\right.\nonumber\\\sbspc&\sbspc&\quad\quad\left.
+\frac{1}{2\Delta{}s^2}\sum_{c}\pi\rho_{c}m_c{R^{(c)}}^3\psi(R^{(c)})
+\sum_{c}\frac{\pi\rho_{c}m_c\varphi_{1}(R^{(c)})\psi(R^{(c)})}{2\Delta{}s}\left(R^{(a)}-R^{(c)}\right)\left(R^{(b)}-R^{(c)}\right)\right]
\nonumber\\
\sbspc&\sbspc&+\frac{e^{-sR^{ab}}}{\Delta{}D}\frac{K^{(a)}}{1-K^{(a)}}\left(1-\delta_{\alpha0}\right)\delta_{\beta0}\left[\frac{1}{s^2}+\frac{R^{ab}}{s}-\frac{R^{(a)}}{\varphi_{0}(R^{(a)})}\left(\frac{1}{s^2}+\frac{R^{(b)}}{2s}\right)-\frac{R^{(a)}}{2\Delta{}s^2}\left(\xi_2-\sum_{c}\pi\rho_{c}m_{c}{R^{(c)}}^2\Theta(R^{(c)})\right)\right.\nonumber\\
\sbspc&\sbspc&\quad\quad+\left.\frac{R^{(a)}R^{(b)}\xi_2}{4\Delta{}s}+\frac{1}{2\Delta{}s^2}\frac{\varphi_{1}(R^{(a)})}{\varphi_{0}(R^{(a)})}s\sum_{c}\pi\rho_{c}m_c{R^{(c)}}^3\psi(R^{(c)})\right.\nonumber\\
\sbspc&\sbspc&\quad\quad+\left.\sum_{c}\frac{\pi\rho_{c}m_c\varphi_{1}(R^{(c)})}{2\Delta{}s}\left(R^{(a)}\Theta(R^{(c)})-\frac{\varphi_{1}(R^{(a)})}{\varphi_{0}(R^{(a)})}s\psi(R^{(c)})R^{(c)}\right)\left(R^{(b)}-R^{(c)}\right)\right]\nonumber\\
\sbspc&\sbspc&+\frac{e^{-sR^{ab}}}{\Delta{}D}\frac{K^{(b)}}{1-K^{(b)}}\delta_{\alpha0}\left(1-\delta_{\beta0}\right)\left[\frac{1}{s^2}+\frac{R^{ab}}{s}-\frac{R^{(b)}}{\varphi_{0}(R^{(b)})}\left(\frac{1}{s^2}+\frac{R^{(a)}}{2s}\right)-\frac{R^{(b)}}{2\Delta{}s^2}\left(\xi_2-\sum_{c}\pi\rho_{c}m_{c}{R^{(c)}}^2\Theta(R^{(c)})\right)\right.\nonumber\\
\sbspc&\sbspc&\quad\quad+\left.\frac{R^{(a)}R^{(b)}\xi_2}{4\Delta{}s}+\frac{1}{2\Delta{}s^2}\frac{\varphi_{1}(R^{(b)})}{\varphi_{0}(R^{(b)})}s\sum_{c}\pi\rho_{c}m_c{R^{(c)}}^3\psi(R^{(c)})\right.\nonumber\\
\sbspc&\sbspc&\quad\quad+\left.\sum_{c}\frac{\pi\rho_{c}m_c\varphi_{1}(R^{(c)})}{2\Delta{}s}\left(R^{(b)}\Theta(R^{(c)})-\frac{\varphi_{1}(R^{(b)})}{\varphi_{0}(R^{(b)})}s\psi(R^{(c)})R^{(c)}\right)\left(R^{(a)}-R^{(c)}\right)\right]\nonumber
\eea
\bea
\sbspc&\sbspc&+\frac{e^{-sR^{ab}}}{\Delta{}D}\frac{K^{(a)}}{1-K^{(a)}}\frac{K^{(b)}}{1-K^{(b)}}\left(1-\delta_{\alpha0}\right)\left(1-\delta_{\beta0}\right)\left[\frac{\varphi_{1}(R^{(a)})}{\varphi_{0}(R^{(a)})}\frac{\varphi_{1}(R^{(b)})}{\varphi_{0}(R^{(b)})}+\left(\frac{R^{(b)}}{2}\frac{\varphi_{1}(R^{(a)})}{\varphi_{0}(R^{(a)})}+\frac{R^{(a)}}{2}\frac{\varphi_{1}(R^{(b)})}{\varphi_{0}(R^{(b)})}\right)\right.\nonumber\\
\sbspc&\sbspc&\quad\quad+\left.\frac{R^{(a)}R^{(b)}\xi_2}{4\Delta{}s}+\frac{1}{2\Delta{}}\frac{\varphi_{1}(R^{(a)})}{\varphi_{0}(R^{(a)})}\frac{\varphi_{1}(R^{(b)})}{\varphi_{0}(R^{(b)})}\sum_{c}\pi\rho_{c}m_c{R^{(c)}}^3\psi(R^{(c)})\right.\nonumber\\
\sbspc&\sbspc&\quad\quad+\left.\frac{s}{2\Delta{}}\frac{\varphi_{1}(R^{(a)})}{\varphi_{0}(R^{(a)})}\frac{\varphi_{1}(R^{(b)})}{\varphi_{0}(R^{(b)})}\sum_{c}\pi\rho_{c}m_c\varphi_{1}(R^{(c)}){R^{(c)}}^2\psi(R^{(c)})+\frac{R^{(a)}R^{(b)}}{2\Delta{}s}\sum_{c}\pi\rho_{c}m_c\varphi_{1}(R^{(c)})\Theta(R^{(c)})\right.\nonumber\\
\sbspc&\sbspc&\quad\quad-\left.\frac{1}{2\Delta}\left(R^{(b)}\frac{\varphi_{1}(R^{(a)})}{\varphi_{0}(R^{(a)})}+R^{(a)}\frac{\varphi_{1}(R^{(b)})}{\varphi_{0}(R^{(b)})}\right)\sum_{c}\pi\rho_{c}m_c\varphi_{1}(R^{(c)})R^{(c)}\psi(R^{(c)})\right]\nonumber\\
\sbspc&\sbspc&+e^{-sR^{(a)}}\frac{\delta_{ab}}{2\pi\rho_{a}m_a}\frac{K^{(a)}}{1-K^{(a)}}\frac{1}{\varphi_{0}(R^{(a)})}\left(\delta_{\alpha{}A}\delta_{\beta{}B}+\delta_{\alpha{}B}\delta_{\beta{}A}\right),
\label{laplaceIm}
\eea
where
\bea
K^{(a)}\sbspc&=\sbspc&\left(\mathbf{\hat{g}}^{(a)}\mathbf{\hat{h}}^{(a)}\right)=\left(\mathbf{\hat{e}}^{(a)}\mathbf{\hat{f}}^{(a)}\right)=\frac{m_a-1}{2m_{a}R^{(a)}}\varphi_{0}(R^{(a)}),\\
\psi(R^{(a)})\sbspc&=\sbspc& 1+\frac{2K^{(a)}}{1-K^{(a)}},\\
\Theta(R^{(a)})\sbspc&=\sbspc&
1+\frac{2K^{(a)}}{1-K^{(a)}}\frac{\varphi_{1}(R^{(a)})}{\varphi_{0}(R^{(a)})}s,
\eea
and
\bea
D\sbspc&=\sbspc&1-\frac{2}{\Delta}\sum_{t}\pi\rho_{t}m_t\varphi_{2}(R^{(t)})\Theta(R^{(t)})\left\{1+\frac{1}{2\Delta}\sum_{c}\pi\rho_cm_c{R^{(c)}}^3\psi(R^{(c)})\right\}\nonumber\\
\sbspc&\sbspc&-\frac{1}{\Delta^2}\sum_{t}\pi\rho_{t}m_t\varphi_{2}(R^{(t)})R^{(t)}\psi(R^{(t)})\left\{\xi_2-\sum_{c}\pi\rho_cm_c{R^{(c)}}^2\Theta(R^{(c)})\right\}\nonumber\\
\sbspc&\sbspc&-\frac{2}{\Delta}\sum_{c}\pi\rho_{c}m_c\varphi_{1}(R^{(c)})\left\{R^{(c)}\left(\psi(R^{(c)})\left[\frac{1}{2}+\frac{\xi_2R^{(c)}}{4\Delta}\right]+\frac{\Theta(R^{(c)})}{2}\right)\right.\\
\sbspc&\sbspc&+\left.\frac{1}{4\Delta}\psi(R^{(c)})\sum_{c}\pi\rho_cm_c\varphi_{1}(R^{(t)})\Theta(R^{(c)})\left(R^{(c)}-R^{(t)}\right)^2\right\}\nonumber.
\eea

Presented above Laplace transform of partial RDF is reperesented as a sum
of five terms that are multiplied by delta symbols and thus some of
them vanish for a ceratin values of the monomer state indices $\alpha$ and
$\beta$. Laplace image of overall radial distribution function can
be calculated by
\bea
G^{ab}_{overall}(s)=\sum_{\alpha\beta}G^{ab}_{\alpha\beta}(s).
\eea
The first four terms of (\ref{laplaceIm}) contribute to
intermolecular RDF, while the last term describes
intramolecular RDF and contains Laplace transform of the delta-function.
The intermolecular and intramolecular parts of the overall RDFs are
related to partial RDFs by
\bea
G^{ab}_{inter}(s)\sbspc&=\sbspc&G^{ab}_{00}(s)+2G^{ab}_{A0}(s)+2G^{ab}_{0A}(s)+4G^{ab}_{AA}(s),\\
G^{ab}_{intra}(s)\sbspc&=\sbspc&2\left(G^{ab}_{AB}(s)-G^{ab}_{AA}(s)\right).
\eea

\section{Thermodynamic properties of the hard-sphere Yukawa chain mixture}

As in our previous paper \cite{Kalyuzhnyi2006}, we use the HTA
to obtain analytical expression for the Helmholtz free energy.
Similarly as in the SAFT\cite{Alejandro1997,Kalyuzhnyi2004} we
assume that free energy per unit volume $f=A/V$ can be written as
following
\be
f=f_{id}+f^{ex}_{HS}+f_{HTA}+f_{chain},
\label{freeenergy1}
\ee
where $f_{id}$ is the ideal gas contribution
\be
\beta{}f_{id}=\sum_{c}\rho_{c}m_{c}\left[{\mathrm{ln}}\left(\Lambda_c^3\rho_c\right)-1\right],
\ee
with $\Lambda_c$ being thermal de Broglie wavelength,
$f^{ex}_{HS}$ is hard sphere excess free energy  and $f_{HTA}$ is
key HTA term that accounts for attractive interaction between
the chain monomers and contains Yukawa potential averaged with respect to the
structure of the reference system. Finally
the third term in (\ref{freeenergy1}) contains contribution from the chain
formation
\be
\beta{}f_{chain}=\sum_{c}\rho_{c}\left(m_{c}-1\right)\left[-{\mathrm{ln}}\left(4\pi\rho_c{R^{(c)}}^2Y\right)+1\right],
\ee
where $Y$ is hard-sphere cavity correlation function.

The hard sphere excess free energy $f^{ex}_{HS}$ and cavity
correlation function $Y$ are taken in
Mansoori-Carnahan-Starling-Leland
approximation\cite{Mansoori1971}, which for hard sphere system of
equal sizes are given by
\bea
\beta{}f_{HS}^{ex}\sbspc&=\sbspc& \frac{1}{2\pi\Delta}l^2R^3\left(1 +
\frac{1}{3\Delta}\right),\\
Y\sbspc&=\sbspc&\frac{1}{\Delta}+\frac{lR^3}{4\Delta^2}+\frac{l^2R^6}{72\Delta^3},
\eea
where $R$ is the hard sphere diameter, $\Delta=1 - {lR^3}/{6}$ and
$l=\pi\sum_{c}\rho^{c}m_c$ is the moment of the chain number densities.

Key HTA term in (\ref{freeenergy1}) holds Yukawa attraction energy
averaged with respect to the overall RDF of reference system
\bea
\beta{}f_{HTA}\sbspc&=\sbspc&
\beta\frac{2}{\pi}\sum_{a}\pi\rho_{a}m_a\sum_{b}\pi\rho_{b}m_b\int_{0}^{\infty}drr^2u^{ab}_{Y}(r)g^{ab}_{overall}(r).
\label{f_hta}
\eea
Due to the choice of interaction potential (\ref{Yukawa})
integral in (\ref{f_hta}) can be substituted by overall RDFs
Laplace transform of chain fluid, hence above formula can be
written as follows
\bea
\beta{}f_{HTA}\sbspc&=\sbspc&
-\frac{2\beta\epsilon_0r_0e^{z^{(0)}R}}{\pi{}}\sum_{a}\pi\rho_{a}m_a\sum_{b}\pi\rho_{b}m_b
e^{-z^{(0)}R}G^{ab}_{overall}(z^{(0)}), \label{f_hta2}
\eea
where
\be
G^{ab}_{overall}(z^{(0)})=\int_{0}^{\infty}dr r
e^{-z^{(0)}r}g^{ab}_{overall}(r).
\ee

We will use obtained expression for the Laplace
transform of the overall RDF
(\ref{laplaceIm}) for HTA free energy contribution (\ref{f_hta2}),
which now reads

\bea
\beta{}f_{HTA}\sbspc&=\sbspc&-\frac{2\beta\epsilon_0r_0}{\pi{}\Delta D}\left\{l^2\left[\frac{1}{{z^{(0)}}^2}+\frac{R}{{z^{(0)}}}-\frac{R^3}{\Delta{}{z^{(0)}}}\frac{\varphi_1(R)}{\varphi_0(R)}v+\frac{lR^4}{4\Delta{}{z^{(0)}}}+\frac{R^3}{2\Delta{}{z^{(0)}}^2}\left(l+2v\right)\right]\right.\\ \label{f_hta3}
\sbspc&\sbspc&\left.+4lv\left[\frac{1}{{z^{(0)}}^2}+\frac{R}{{z^{(0)}}}-\frac{R}{\varphi_0(R)}\left(\frac{1}{{z^{(0)}}^2}+\frac{R}{2{z^{(0)}}}\right)-\frac{R^3}{\Delta{}{z^{(0)}}}\frac{\varphi_1(R)}{\varphi_0(R)}v+\frac{lR^4}{4\Delta{}{z^{(0)}}}+\frac{\left(l+2v\right)R^3}{2\Delta{}{z^{(0)}}}\frac{\varphi_1(R)}{\varphi_0(R)}\right]\right.\nonumber\\
\sbspc&\sbspc&\left.+4v^2\left[\left(\frac{\varphi_1(R)}{\varphi_0(R)}\right)^2+R\frac{\varphi_1(R)}{\varphi_0(R)}+\frac{lR^4}{4\Delta{}{z^{(0)}}}+\frac{l+2v}{\Delta}\left(\left(\frac{R^3}{2}+\frac{\varphi_1(R){z^{(0)}}R^2}{2}\right)\left(\frac{\varphi_1(R)}{\varphi_0(R)}\right)^2-R^2\varphi_1(R)\frac{\varphi_1(R)}{\varphi_0(R)}\right)\right.\right.\nonumber\\
\sbspc&\sbspc&\left.\left.+\frac{\left(l+2v\frac{\varphi_1(R)}{\varphi_0(R)}{z^{(0)}}\right)R^2\varphi_1(R)}{2\Delta{}{z^{(0)}}}\right]\right\}
-\frac{2\beta\epsilon_0r_0}{\pi{}}
\frac{2v}{\varphi_0(R)}\nonumber
\eea
where
\bea
D\sbspc&=\sbspc&1-\frac{2}{\Delta}\varphi_2(R)\left(l+2v\frac{\varphi_1(R)}{\varphi_0(R)}s\right)\left(1+\frac{R^3}{2\Delta}\left(l+2v\right)\right)
+
\frac{R^3\varphi_2(R)}{\Delta^2}\left(l+2v\right)2v\frac{\varphi_1(R)}{\varphi_0(R)}s\\
\sbspc&\sbspc&-\frac{1}{\Delta}\varphi_1(R)R
\left[\left(l+2v\right)\left(1+\frac{lR^3}{2\Delta}\right)+\left(l+2v\frac{\varphi_1(R)}{\varphi_0(R)}s\right)\right],\nonumber
\eea
\bea
l\sbspc&=\sbspc&\pi\sum_{c}\rho^{c}m_c,\\
v\sbspc&=\sbspc&\pi\sum_{c}\rho^{c}m_c\frac{K_c}{1-K_c},
\eea
are the moments of the chain species distribution function,
and
\bea
\varphi_{2}(R^{(b)})\sbspc&=\sbspc&\frac{1-{z^{(0)}}R^{(b)}+{z^{(0)}}^2{R^{(b)}}^2/2-e^{-{z^{(0)}}R^{(b)}}}{{z^{(0)}}^3},\\
\varphi_{1}(R^{(b)})\sbspc&=\sbspc&\frac{1-{z^{(0)}}R^{(b)}-e^{-{z^{(0)}}R^{(b)}}}{{z^{(0)}}^2},\\
\varphi_{0}(R^{(b)})\sbspc&=\sbspc&\frac{1-e^{-{z^{(0)}}R^{(b)}}}{{z^{(0)}}}.
\eea

To study the phase behavior, we will also need expressions for
the pressure of the system and chemical potential of the chain molecules, which are
derived from the above expression for Helmholtz free energy
(\ref{freeenergy1})-(\ref{f_hta3})
\bea
\beta{}P\sbspc&=\sbspc&\beta{}\sum_c\rho_{c}\mu_c-\beta{}f=\beta{}P_{HS}+\beta{}P_{chain}+\beta{}P_{HTA},\\
\beta\mu_c\sbspc&=\sbspc&\frac{\partial\left(\beta{}f\right)}{\partial{}\rho_c}=\beta\mu_{c,id}+\beta\mu^{ex}_{c,HS}+\beta\mu_{c,chain}+\beta\mu_{c,HTA}.
\eea
Here $\mu_{c,id}$, $\mu^{ex}_{c,HS}$, $\mu_{c,chain}$,
$\mu_{c,HTA}$ and $P_{chain}$, $P_{HTA}$ hold contributions from
corresponding free energies $f_{id}$, $f^{ex}_{HS}$, $f_{chain}$
and $f_{HTA}$, except for $\beta{}P_{HS}$ that holds sum of
contributions from $f_{id}$ and $f^{ex}_{HS}$. These expressions
for excess chemical potential and pressure can be formulated in
terms of moments and their density derivatives
\bea
\beta\mu_{c,id}\sbspc&=\sbspc&m_c\mathrm{ln}\left\{\Lambda^3_c\rho_c\right\}\\
\beta\mu_{c,HS}^{ex}\sbspc&=\sbspc&\frac{lR^3l(c)}{\pi\Delta}\left(1+\frac{1}{3\Delta}\right)+\frac{l^2R^6l(c)}{12\pi\Delta^2}\left(1+\frac{2}{3\Delta}\right)\\
\beta\mu_{c,chain}\sbspc&=\sbspc&-(m_c-1)\mathrm{ln}\left\{4\pi\rho_cR^2Y\right\}-\frac{\left(l-\pi\rho\right)}{Y}\frac{\partial{}Y}{\partial\rho_c}\\
\beta\mu_{c,HTA}\sbspc&=\sbspc&-\frac{4\beta\epsilon_0r_0e^{z^{(0)}R}}{\pi{}}\pi\rho_{c}m_c\sum_{a}\pi\rho_{a}m_a
G^{ac}_{overall}(z^{(0)})\\
\sbspc&\sbspc&-\frac{2\beta\epsilon_0r_0e^{z^{(0)}R}}{\pi{}}\sum_{a}\pi\rho_{a}m_a\sum_{b}\pi\rho_{b}m_b
\frac{\partial{}G^{ab}_{overall}(z^{(0)})}{\partial\rho_c}\nonumber\\
\beta{}P_{HS}\sbspc&=\sbspc&\frac{l}{\Delta\pi}\left(1+\frac{lR^3}{2\Delta}+\frac{l^2R^6}{12\Delta^2}+\frac{l^3R^9}{216\Delta^2}\right)\\
\beta{}P_{chain}\sbspc&=\sbspc&\frac{l-\pi\rho}{\pi}\left(\frac{1}{Y}\left[\frac{5lR^3}{12\Delta^2}+\frac{l^2R^6}{12\Delta^3}\right]+1\right)\\
\beta{}P_{HTA}\sbspc&=\sbspc&-\frac{2\beta\epsilon_0r_0e^{z^{(0)}R}}{\pi{}}\sum_{a}\pi\rho_{a}m_a\sum_{b}\pi\rho_{b}m_b
G^{ac}_{overall}(z^{(0)})\\
\sbspc&\sbspc&-\frac{2\beta\epsilon_0r_0e^{z^{(0)}R}}{\pi{}}\sum_{a}\pi\rho_{a}m_a\sum_{b}\pi\rho_{b}m_b
\sum_{c}\rho_{c}\frac{\partial{}G^{ab}_{overall}(z^{(0)})}{\partial\rho_c}\nonumber
\eea
where
\bea
\frac{\partial{}Y}{\partial\rho_c}\sbspc&=\sbspc&\frac{5R^3l_c}{12\Delta^3}+\frac{lR^6l_c}{9\Delta^2}+\frac{l^3R^9l_c}{144\Delta^4},\\
l_c\sbspc&=\sbspc&\frac{\partial{}l}{\partial\rho_c}=\pi{}m_c,\\
v_c\sbspc&=\sbspc&\frac{\partial{}v}{\partial\rho_c}=\pi{}m_c\frac{K_c}{1-K_c}.
\eea
and $\rho=\sum_c\rho_c$ is total number density. Because of the
size expressions for $\mu_{c,HTA}$ and $P_{HTA}$ via the moments
are shown in the Appendix.

Thus, we can conclude, that our model belongs to the family of the TFE models,
since its thermodynamic properties can be
represented by finite number of (generalized) density distribution
moments.

\section{Two phase equilibria}

When at constant $T$ mother phase with the number density
$\rho_c^{(0)}$ of the chain species $c$ and volume $V^{(0)}$ separates
into equilibrium daughter phases, then all the chemical potentials of
every chain species and all pressures in each of the daughter phases
are equal. Thus for the two-phase equilibrium we have
\bea
\mu^{(1)}_c\left(\left\{\rho_c^{(1)}\right\}\right)\sbspc&=\sbspc&\mu^{(2)}_c\left(\left\{\rho_c^{(2)}\right\}\right), \quad for \; every \; c \label{equilibr1}\\
P^{(1)}\sbspc&=\sbspc&P^{(2)},\nonumber
\eea
where 
$P^{(p)}$ and
$\mu^{(p)}_c\left(\left\{\rho_c^{(p)}\right\}\right)$ are pressure
and chemical potential of phase $p$. We have explicitly specified
the dependence on the species number densities
$\left\{\rho_c^{(p)}\right\}$ to indicate that they differ between
phases.
In addition particle number
 and volume conservation rules have to be satisfied
\bea
f^{(0)}_c\sbspc&=\sbspc&x^{(1)}f^{(1)}_c+x^{(2)}f^{(2)}_c \label{particle_conserv}\\
v^{(0)}\sbspc&=\sbspc&x^{(1)}v^{(1)}+x^{(2)}v^{(2)}\label{volume_conserv},
\eea
together with normalization condition
\bea
1=x^{(1)}+x^{(2)}\label{normalization1},
\eea
where $x^{(p)}=N^{(p)}/N^{(0)}$ for $p=1,2$ is the chain molecules
number fractions of phase $p$, $N^{(p)}=V^{(p)}\rho^{(p)}$ is the
total chain molecule number in phase $p$, and
$f^{(p)}_c=\rho^{(p)}_c/\rho^{(p)}=N^{(p)}_c/N^{(p)}$ is molecular
species distribution function
\bea
\sum_{c}f^{(p)}_c=1.\label{distr_norm}
\eea


Assuming that we have the number of the chain species $M$, conditions
(\ref{equilibr1})-(\ref{distr_norm}) form closed set of $M+4$
equations for $M+4$ unknowns: $f_c^{(1)}$, $\rho^{(1)}$,
$\rho^{(2)}$, $x^{(1)}$, $x^{(2)}$. As in our previous work
\cite{Kalyuzhnyi2006}, to solve above equations, we will employ
general scheme developed by Bellier-Castella et
al\cite{Bellier-Castella2000}. The scheme is 
suitable
only for truncated free energy models where very
efficient transition to new unknown variables, represented by
the moments, can be done. According to the scheme, the set of $M+4$
equations, applied to our model, transforms to $4$ equations in
two unknown moments of the first phase $l^{(1)}$, $v^{(1)}$ and two
unknown densities of the daughter phases $\rho^{(1)}$, $\rho^{(2)}$:
\bea
\sbspc&\sbspc&\rho^{(1)}=\rho^{(1)}\sum_{c}{f^{(0)}_{c}H_c\left(T;\rho^{(1)},l^{(1)}v^{(1)};\rho^{(2)},l^{(2)}v^{(2)}\right)},\label{phs_eq_moment_eq1}\\
\sbspc&\sbspc&l^{(1)}=\rho^{(1)}\sum_{c}{l_{c}f^{(0)}_{c}H_c\left(T;\rho^{(1)},l^{(1)}v^{(1)};\rho^{(2)},l^{(2)}v^{(2)}\right)},\\
\sbspc&\sbspc&v^{(1)}=\rho^{(1)}\sum_{c}{v_{c}f^{(0)}_{c}H_c\left(T;\rho^{(1)},l^{(1)}v^{(1)};\rho^{(2)},l^{(2)}v^{(2)}\right)}\label{phs_eq_moment_eq2},\\
\sbspc&\sbspc&P^{(1)}\left(T;\rho^{(1)},l^{(1)}v^{(1)}\right)=P^{(2)}\left(T;\rho^{(2)},l^{(2)}v^{(2)}\right),
\eea
where
\bea
H_c\left(T;\rho^{(1)},l^{(1)}v^{(1)};\rho^{(2)},l^{(2)}v^{(2)}\right)\sbspc&=\sbspc&
\frac{\left(\rho^{(1)}-\rho^{(2)}\right)A^{12}_c\left(T;\rho^{(1)},l^{(1)}v^{(1)};\rho^{(2)},l^{(2)}v^{(2)}\right)}{\left(
\frac{\rho^{(2)}\rho^{(1)}}{\rho^{(0)}}-\rho^{(2)}\right)+\left(\rho^{(1)}-\frac{\rho^{(2)}\rho^{(1)}}{\rho^{(0)}}\right)A^{12}_c\left(T;\rho^{(1)},l^{(1)}v^{(1)};\rho^{(2)},l^{(2)}v^{(2)}\right)},\\
A^{12}_c\left(T;\rho^{(1)},l^{(1)}v^{(1)};\rho^{(2)},l^{(2)}v^{(2)}\right)\sbspc&=\sbspc&\frac{\rho^{(2)}}{\rho^{(1)}}
\exp\left[\mu_{c}^{ex(2)}\left(T;\rho^{(2)},l^{(2)}v^{(2)}\right)-\mu^{ex(1)}_{c}\left(T;\rho^{(1)},l^{(1)}v^{(1)}\right)\right],
\eea
and $\mu^{ex}_c=\mu_c-\mu_{id,c}$ is excess chemical potential.
Note that two moments of the second phase $l^{(2)}$ and $v^{(2)}$
aren't unknown variables, because they can be found from the relation
between moments of the phase $0$, which are known in advance, and
moments of the phase $1$
\be
m^{(2)}=\frac{\rho^{(1)}-\rho^{(2)}}{\rho^{(1)}-\rho^{(0)}}m^{(0)}
+\frac{\rho^{(2)}-\rho^{(0)}}{\rho^{(1)}-\rho^{(0)}}m^{(1)},
\ee
where $m$ is one of the moments $l$ or $v$. Set of equations presented above
are used to locate the binodals. To obtain cloud
and shadow curves, we consider the limit
$\rho^{(2)}\rightarrow\rho^{(0)}$, which employs that second
(cloud) phase is very close to parent phase and thus there is only
an infinitesimal amount of the first (shadow) phase . In such case it
can be easily shown that
\bea
\lim_{\rho^{(2)}\rightarrow\rho^{(0)}}H_c\left(T;\rho^{(1)},l^{(1)}v^{(1)};\rho^{(2)},l^{(2)}v^{(2)}\right)\sbspc&=\sbspc&
A^{12}_c\left(T;\rho^{(1)},l^{(1)}v^{(1)};\rho^{(2)},l^{(2)}v^{(2)}\right),
\eea
and therefore in equations
(\ref{phs_eq_moment_eq1})-(\ref{phs_eq_moment_eq2}), $H_c$ should
be replaced by $A^{12}_c$.

\section{Results and discussion}

We will present now phase diagrams for the two systems that differ
from each other due to the mother distribution function $f^{(0)}_c$. In
contrast to the other studies of polydisperse fluids, we assume
that distribution function depends on discrete species index $c$,
rather than continuous one. But the species number should be large
enough to expose properties of polydisperse systems. Each chain
species $c$ has length $m_c=c$ monomers, for $c=1..M$. In most cases we consider
the systems with $M=100$. For the analysis of the cutoff effects we
present results for $200$ and $300$ chain species.

\begin{figure}[t]
\centerline{
\begin{picture}(0,0)%
\includegraphics{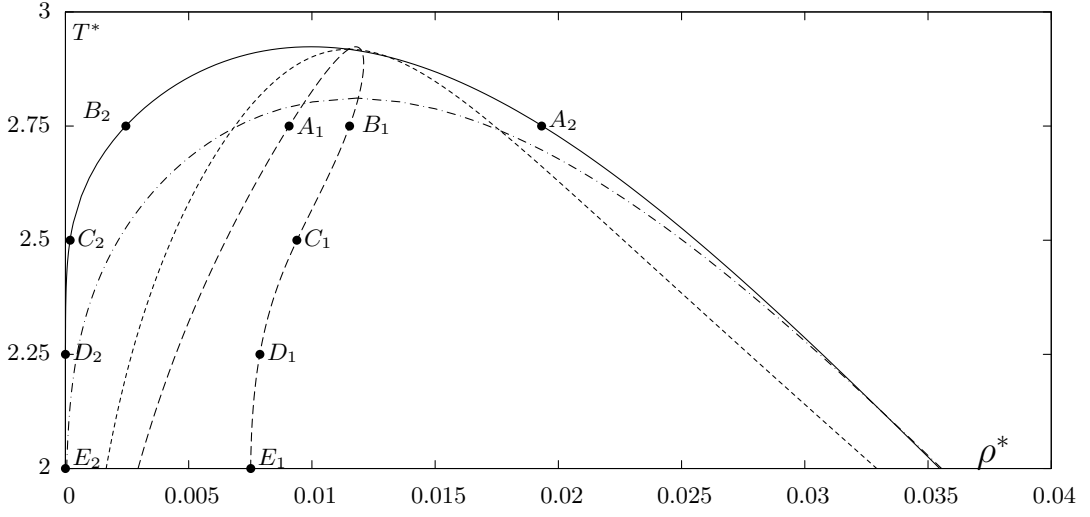}%
\end{picture}%
\begingroup
\setlength{\unitlength}{0.0200bp}%
\begin{picture}(21600,10259)(0,0)%
\put(1925,1100){\makebox(0,0)[r]{\strut{} 2}}%
\put(1925,3253){\makebox(0,0)[r]{\strut{} 2.25}}%
\put(1925,5405){\makebox(0,0)[r]{\strut{} 2.5}}%
\put(1925,7558){\makebox(0,0)[r]{\strut{} 2.75}}%
\put(1925,9710){\makebox(0,0)[r]{\strut{} 3}}%
\put(2200,550){\makebox(0,0){\strut{} 0}}%
\put(4522,550){\makebox(0,0){\strut{} 0.005}}%
\put(6844,550){\makebox(0,0){\strut{} 0.01}}%
\put(9166,550){\makebox(0,0){\strut{} 0.015}}%
\put(11488,550){\makebox(0,0){\strut{} 0.02}}%
\put(13809,550){\makebox(0,0){\strut{} 0.025}}%
\put(16131,550){\makebox(0,0){\strut{} 0.03}}%
\put(18453,550){\makebox(0,0){\strut{} 0.035}}%
\put(20775,550){\makebox(0,0){\strut{} 0.04}}%
\put(6554,7514){\makebox(0,0)[l]{\strut{}$A_1$}}%
\put(11311,7644){\makebox(0,0)[l]{\strut{}$A_2$}}%
\put(7786,7558){\makebox(0,0)[l]{\strut{}$B_1$}}%
\put(2525,7816){\makebox(0,0)[l]{\strut{}$B_2$}}%
\put(6700,5405){\makebox(0,0)[l]{\strut{}$C_1$}}%
\put(2428,5405){\makebox(0,0)[l]{\strut{}$C_2$}}%
\put(6003,3253){\makebox(0,0)[l]{\strut{}$D_1$}}%
\put(2339,3253){\makebox(0,0)[l]{\strut{}$D_2$}}%
\put(5831,1272){\makebox(0,0)[l]{\strut{}$E_1$}}%
\put(2339,1272){\makebox(0,0)[l]{\strut{}$E_2$}}%
\put(2339,9280){\makebox(0,0)[l]{\strut{}$T^*$}}%
\put(19382,1401){\makebox(0,0)[l]{\strut{}\Large $\rho^*$}}%
\end{picture}%
\endgroup
}\footnotesize \caption{Phase diagram ($T^*$ vs $\rho^*$) of
polydisperse chain fluid with $D_m=0.5$ beta distribution
function. Cloud curve, shadow curve are represented by solid and
dashed curves respectively. Dotted curve denotes critical binodal,
and dash-dotted line represents monodisperse binodal. Also five
points on shadow curve $A_1$,$B_1$,$C_1$,$D_1$,$E_1$ together with
corresponding equilibrium points $A_2$,$B_2$,$C_2$,$D_2$,$E_2$ on
cloud curve, mark points for which distribution functions are
presented on \ref{fig9}.}\label{fig1}
\end{figure}

\begin{figure}[t]
\centerline{
\begin{picture}(0,0)%
\includegraphics{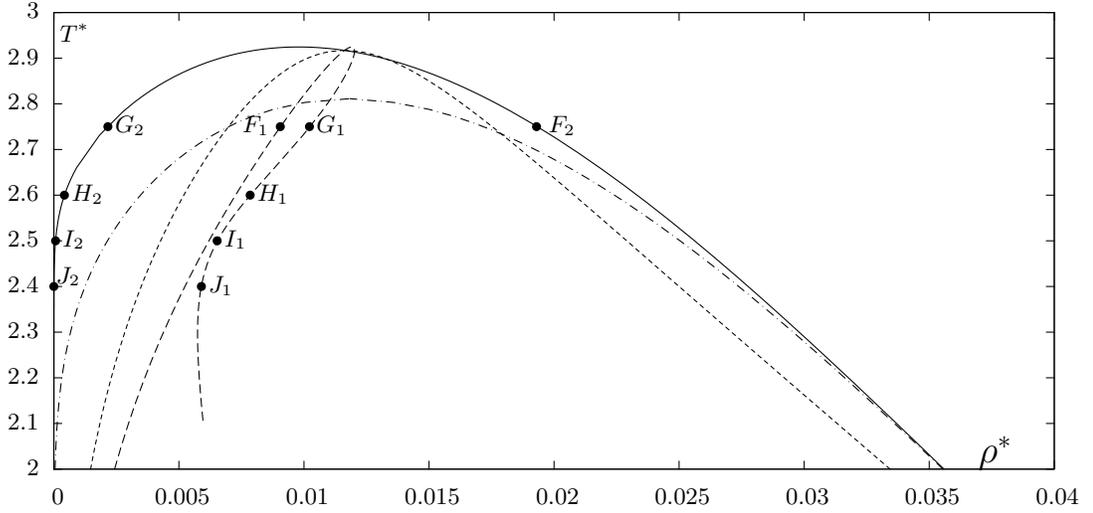}%
\end{picture}%
\begingroup
\setlength{\unitlength}{0.0200bp}%
\begin{picture}(21600,10259)(0,0)%
\put(1650,1100){\makebox(0,0)[r]{\strut{} 2}}%
\put(1650,1961){\makebox(0,0)[r]{\strut{} 2.1}}%
\put(1650,2822){\makebox(0,0)[r]{\strut{} 2.2}}%
\put(1650,3683){\makebox(0,0)[r]{\strut{} 2.3}}%
\put(1650,4544){\makebox(0,0)[r]{\strut{} 2.4}}%
\put(1650,5405){\makebox(0,0)[r]{\strut{} 2.5}}%
\put(1650,6266){\makebox(0,0)[r]{\strut{} 2.6}}%
\put(1650,7127){\makebox(0,0)[r]{\strut{} 2.7}}%
\put(1650,7988){\makebox(0,0)[r]{\strut{} 2.8}}%
\put(1650,8849){\makebox(0,0)[r]{\strut{} 2.9}}%
\put(1650,9710){\makebox(0,0)[r]{\strut{} 3}}%
\put(1925,550){\makebox(0,0){\strut{} 0}}%
\put(4281,550){\makebox(0,0){\strut{} 0.005}}%
\put(6638,550){\makebox(0,0){\strut{} 0.01}}%
\put(8994,550){\makebox(0,0){\strut{} 0.015}}%
\put(11350,550){\makebox(0,0){\strut{} 0.02}}%
\put(13706,550){\makebox(0,0){\strut{} 0.025}}%
\put(16063,550){\makebox(0,0){\strut{} 0.03}}%
\put(18419,550){\makebox(0,0){\strut{} 0.035}}%
\put(20775,550){\makebox(0,0){\strut{} 0.04}}%
\put(5483,7558){\makebox(0,0)[l]{\strut{}$F_1$}}%
\put(11256,7558){\makebox(0,0)[l]{\strut{}$F_2$}}%
\put(6883,7558){\makebox(0,0)[l]{\strut{}$G_1$}}%
\put(3084,7558){\makebox(0,0)[l]{\strut{}$G_2$}}%
\put(5766,6266){\makebox(0,0)[l]{\strut{}$H_1$}}%
\put(2265,6266){\makebox(0,0)[l]{\strut{}$H_2$}}%
\put(5144,5405){\makebox(0,0)[l]{\strut{}$I_1$}}%
\put(2099,5405){\makebox(0,0)[l]{\strut{}$I_2$}}%
\put(4847,4544){\makebox(0,0)[l]{\strut{}$J_1$}}%
\put(1972,4716){\makebox(0,0)[l]{\strut{}$J_2$}}%
\put(2043,9280){\makebox(0,0)[l]{\strut{}$T^*$}}%
\put(19361,1401){\makebox(0,0)[l]{\strut{}\Large $\rho^*$}}%
\end{picture}%
\endgroup
}\footnotesize \caption{The same as figure \ref{fig1} but for
Shultz distribution with $z=1$. Corresponding distributions for
specified points are presented on \ref{fig10}. } \label{fig2}
\end{figure}

In the following we will consider two systems with two
different distribution functions. The first mother distribution
function was chosen to be Beta-distribution
\be
f^{(0)}_c=\frac{1}{B(\alpha,\beta)}\left(\frac{c}{M}\right)^{\alpha-1}\left(1-\frac{c}{M}\right)^{\beta-1},
\ee
for $c=1..M$, where $B(\alpha,\beta)$ is beta function and
$\alpha$ and $\beta$ are related to mean chain length
($m_0=\left\langle{}m\right\rangle{}$) and mean square length
$\left\langle{}m^2\right\rangle{}$ by
\bea
\sbspc&\sbspc&\alpha=\frac{M-m_0\left(1+D_m\right)}{MD_m},\\
\sbspc&\sbspc&\beta=\left(\frac{M-m_0}{m_0}\right)\alpha,\\
\sbspc&\sbspc&D_m=\frac{\left\langle{}m^2\right\rangle}{{m_0}^2}-1=\frac{{\sigma_m}^2}{{m_0}^2},
\eea
with
$\sigma_m=\sqrt{\left\langle{}\left(m-\left\langle{}m\right\rangle\right)^2\right\rangle}$
being standard chain length deviation.

The second distribution function was chosen to be represented by the Shultz distribution
\be
f^{(0)}_c=\frac{1}{z!}\left(\frac{z+1}{m_0}\right)^{z+1}c^{z}\exp{\left[-\left(\frac{z+1}{m_0}\right)c\right]},
\ee
for $c=1..M$, where $z$ is dispersity controlling parameter.

In both cases the dispersity of the distribution functions were chosen
to be equal, with standard deviation $\sigma_m = 11.31$ that
corresponds to $D_m=0.5$ for beta distribution and $z=1$ for
Shultz distribution, and with average chain length $m_0=16$ for
both distributions.

\begin{figure}[t]
\centerline{
\begin{picture}(0,0)%
\includegraphics{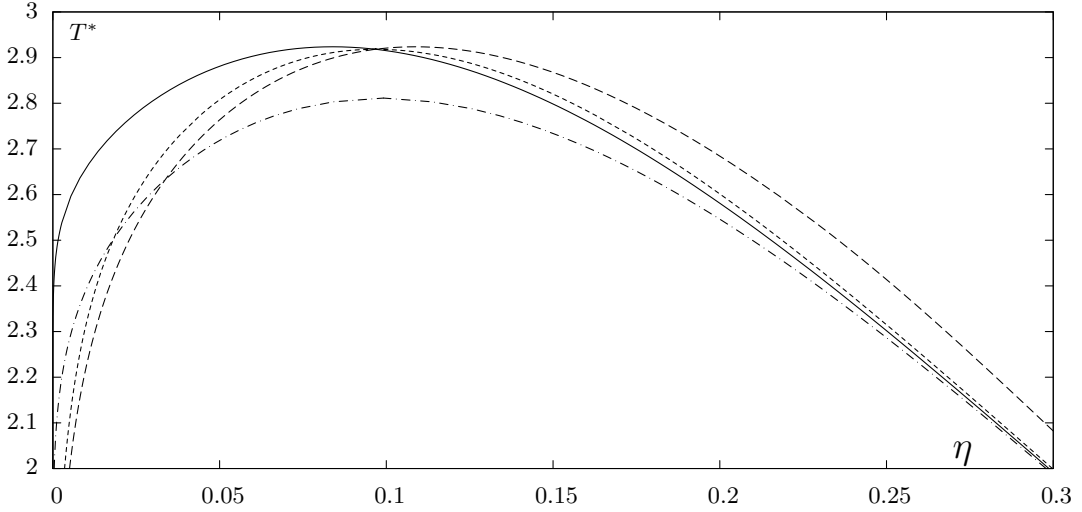}%
\end{picture}%
\begingroup
\setlength{\unitlength}{0.0200bp}%
\begin{picture}(21600,10259)(0,0)%
\put(1650,1100){\makebox(0,0)[r]{\strut{} 2}}%
\put(1650,1961){\makebox(0,0)[r]{\strut{} 2.1}}%
\put(1650,2822){\makebox(0,0)[r]{\strut{} 2.2}}%
\put(1650,3683){\makebox(0,0)[r]{\strut{} 2.3}}%
\put(1650,4544){\makebox(0,0)[r]{\strut{} 2.4}}%
\put(1650,5405){\makebox(0,0)[r]{\strut{} 2.5}}%
\put(1650,6266){\makebox(0,0)[r]{\strut{} 2.6}}%
\put(1650,7127){\makebox(0,0)[r]{\strut{} 2.7}}%
\put(1650,7988){\makebox(0,0)[r]{\strut{} 2.8}}%
\put(1650,8849){\makebox(0,0)[r]{\strut{} 2.9}}%
\put(1650,9710){\makebox(0,0)[r]{\strut{} 3}}%
\put(1925,550){\makebox(0,0){\strut{} 0}}%
\put(5067,550){\makebox(0,0){\strut{} 0.05}}%
\put(8208,550){\makebox(0,0){\strut{} 0.1}}%
\put(11350,550){\makebox(0,0){\strut{} 0.15}}%
\put(14492,550){\makebox(0,0){\strut{} 0.2}}%
\put(17633,550){\makebox(0,0){\strut{} 0.25}}%
\put(20775,550){\makebox(0,0){\strut{} 0.3}}%
\put(2239,9280){\makebox(0,0)[l]{\strut{}$T^*$}}%
\put(18890,1401){\makebox(0,0)[l]{\strut{}\Large $\eta$}}%
\end{picture}%
\endgroup
}\footnotesize\caption{Phase diagram ($T^*$ vs
$\eta=(\pi/6)R^3\sum_{c}m_c\rho_c$) of polydisperse chain fluid
with $D_m=0.5$ beta distribution function. The curves denote the
same as in \ref{fig1} and \ref{fig2}.}\label{fig3}
\end{figure}

\begin{figure}[t]
\centerline{
\begin{picture}(0,0)%
\includegraphics{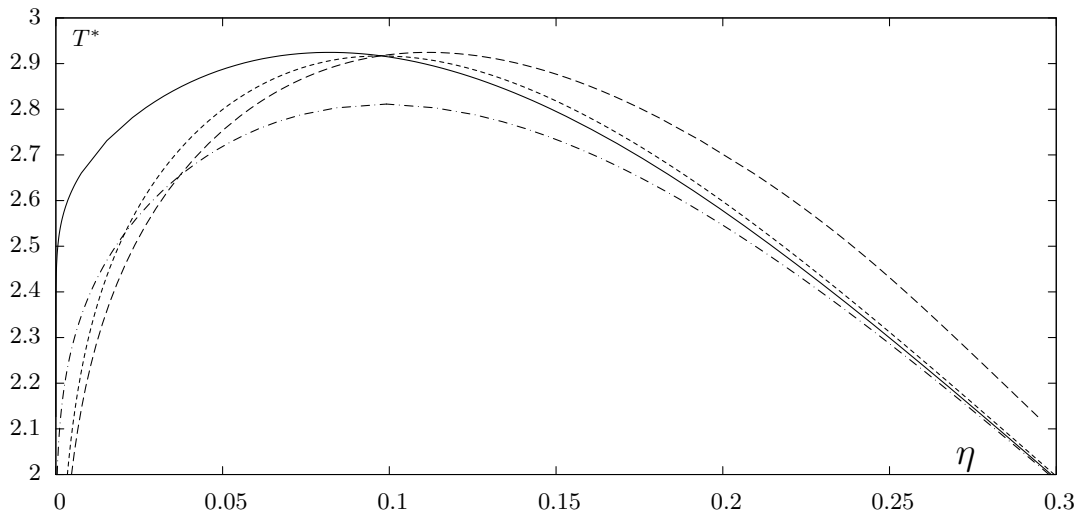}%
\end{picture}%
\begingroup
\setlength{\unitlength}{0.0200bp}%
\begin{picture}(21600,10259)(0,0)%
\put(1650,1100){\makebox(0,0)[r]{\strut{} 2}}%
\put(1650,1961){\makebox(0,0)[r]{\strut{} 2.1}}%
\put(1650,2822){\makebox(0,0)[r]{\strut{} 2.2}}%
\put(1650,3683){\makebox(0,0)[r]{\strut{} 2.3}}%
\put(1650,4544){\makebox(0,0)[r]{\strut{} 2.4}}%
\put(1650,5405){\makebox(0,0)[r]{\strut{} 2.5}}%
\put(1650,6266){\makebox(0,0)[r]{\strut{} 2.6}}%
\put(1650,7127){\makebox(0,0)[r]{\strut{} 2.7}}%
\put(1650,7988){\makebox(0,0)[r]{\strut{} 2.8}}%
\put(1650,8849){\makebox(0,0)[r]{\strut{} 2.9}}%
\put(1650,9710){\makebox(0,0)[r]{\strut{} 3}}%
\put(1925,550){\makebox(0,0){\strut{} 0}}%
\put(5067,550){\makebox(0,0){\strut{} 0.05}}%
\put(8208,550){\makebox(0,0){\strut{} 0.1}}%
\put(11350,550){\makebox(0,0){\strut{} 0.15}}%
\put(14492,550){\makebox(0,0){\strut{} 0.2}}%
\put(17633,550){\makebox(0,0){\strut{} 0.25}}%
\put(20775,550){\makebox(0,0){\strut{} 0.3}}%
\put(2239,9280){\makebox(0,0)[l]{\strut{}$T^*$}}%
\put(18890,1401){\makebox(0,0)[l]{\strut{}\Large $\eta$}}%
\end{picture}%
\endgroup
}\footnotesize \caption{ Same as \ref{fig3}, but for Shultz
distribution with $z=1$.}\label{fig4}
\end{figure}

The corresponding phase diagrams for both systems are demonstrated
on figures \ref{fig1} and \ref{fig2} in terms of the number densities
and on figures \ref{fig3} and \ref{fig4} in terms the volume fraction
$\eta$. Phase diagrams on figures \ref{fig3} and \ref{fig4} are
almost indistinguishable, but 
figures \ref{fig1},\ref{fig2} have one significant difference:
liquid part of the shadow curve for the system with Shultz distribution
has smaller density than that of them phase diagram with Beta
distribution. This difference of the phase behavior can be
explained by inspecting distribution functions on figures
\ref{fig9} and \ref{fig10} for corresponding points marked on
the phase diagrams on figures \ref{fig1} and \ref{fig2}. Liquid shadow curves for
both systems consist of the chains longer than average, in contrast to
the gas shadow curves, that are mainly composed of shorter chains.
As one can see for the system with Shultz distribution the chains in liquid
phase are on average larger than for the systems with Beta distribution. The
origin of these effects lies in the difference of the tails for the mother
phase distributions (either Beta or Shultz). For Shultz distribution the tail is
"fatter" than for Beta distribution. Being more precise at
cutoff distance $m_{cutoff}=M=100$ the Shultz distribution tends
to finite value as opposed to Beta whose value tends to 0. These
differences in compositions of the two liquid shadow phases are
the reason for the discrepancy between phase diagrams. Note, that
volume fractions are close, only average chain lengths and as a
result number densities differ.

\begin{figure}[t]
\centerline{
\begin{picture}(0,0)%
\includegraphics{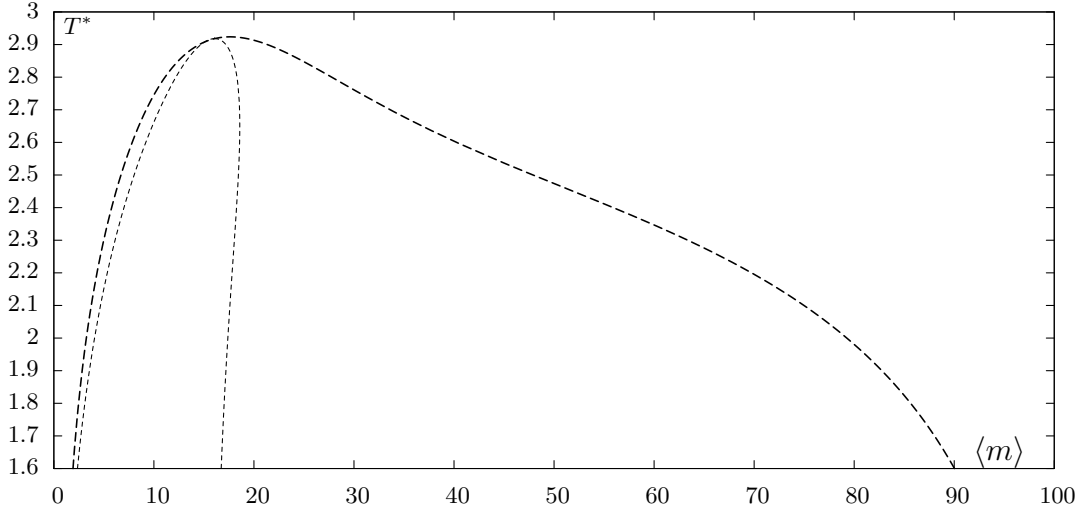}%
\end{picture}%
\begingroup
\setlength{\unitlength}{0.0200bp}%
\begin{picture}(21600,10259)(0,0)%
\put(1650,1100){\makebox(0,0)[r]{\strut{} 1.6}}%
\put(1650,1715){\makebox(0,0)[r]{\strut{} 1.7}}%
\put(1650,2330){\makebox(0,0)[r]{\strut{} 1.8}}%
\put(1650,2945){\makebox(0,0)[r]{\strut{} 1.9}}%
\put(1650,3560){\makebox(0,0)[r]{\strut{} 2}}%
\put(1650,4175){\makebox(0,0)[r]{\strut{} 2.1}}%
\put(1650,4790){\makebox(0,0)[r]{\strut{} 2.2}}%
\put(1650,5405){\makebox(0,0)[r]{\strut{} 2.3}}%
\put(1650,6020){\makebox(0,0)[r]{\strut{} 2.4}}%
\put(1650,6635){\makebox(0,0)[r]{\strut{} 2.5}}%
\put(1650,7250){\makebox(0,0)[r]{\strut{} 2.6}}%
\put(1650,7865){\makebox(0,0)[r]{\strut{} 2.7}}%
\put(1650,8480){\makebox(0,0)[r]{\strut{} 2.8}}%
\put(1650,9095){\makebox(0,0)[r]{\strut{} 2.9}}%
\put(1650,9710){\makebox(0,0)[r]{\strut{} 3}}%
\put(1925,550){\makebox(0,0){\strut{} 0}}%
\put(3810,550){\makebox(0,0){\strut{} 10}}%
\put(5695,550){\makebox(0,0){\strut{} 20}}%
\put(7580,550){\makebox(0,0){\strut{} 30}}%
\put(9465,550){\makebox(0,0){\strut{} 40}}%
\put(11350,550){\makebox(0,0){\strut{} 50}}%
\put(13235,550){\makebox(0,0){\strut{} 60}}%
\put(15120,550){\makebox(0,0){\strut{} 70}}%
\put(17005,550){\makebox(0,0){\strut{} 80}}%
\put(18890,550){\makebox(0,0){\strut{} 90}}%
\put(20775,550){\makebox(0,0){\strut{} 100}}%
\put(2114,9403){\makebox(0,0)[l]{\strut{}$T^*$}}%
\put(19267,1407){\makebox(0,0)[l]{\strut{}\large $\left\langle{}m\right\rangle$}}%
\end{picture}%
\endgroup
}\footnotesize \caption{Shows shadow and critical binodal curves
for system with Beta distribution, denoted respectively by dashed
and dotted lines, in terms if $T^*$ and phase average chain length
$\left\langle{}m\right\rangle$. }\label{fig5}
\end{figure}

\begin{figure}[t]
\centerline{
\begin{picture}(0,0)%
\includegraphics{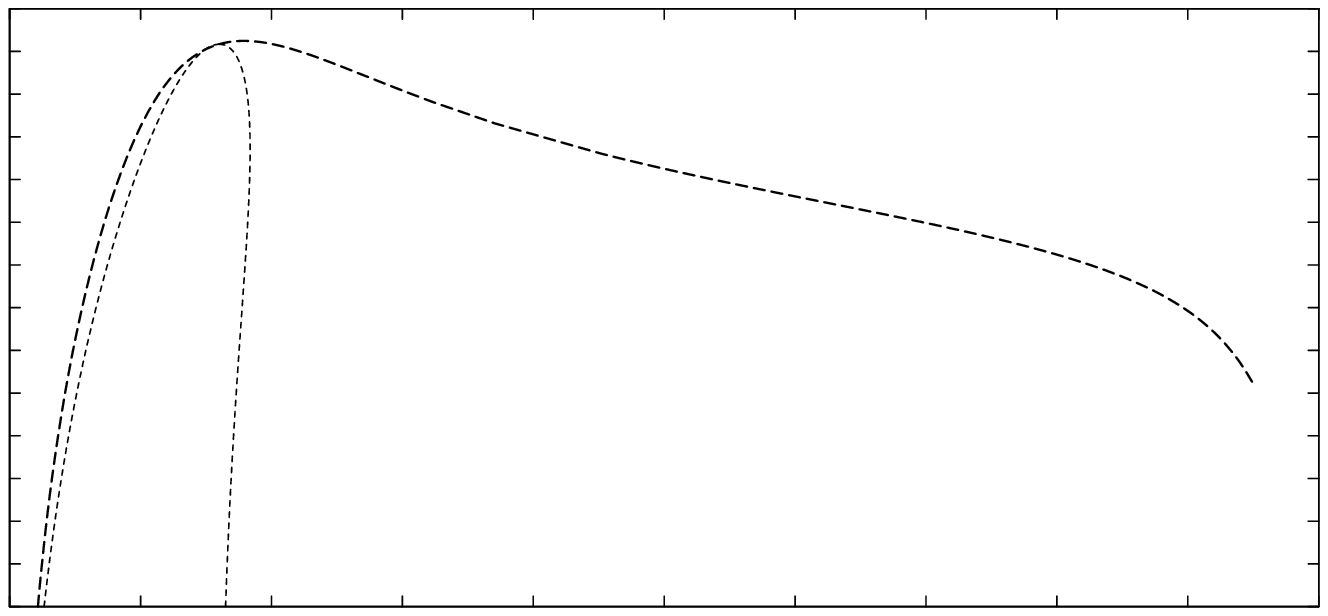}%
\end{picture}%
\begingroup
\setlength{\unitlength}{0.0200bp}%
\begin{picture}(21600,10259)(0,0)%
\put(1650,1100){\makebox(0,0)[r]{\strut{} 1.6}}%
\put(1650,1715){\makebox(0,0)[r]{\strut{} 1.7}}%
\put(1650,2330){\makebox(0,0)[r]{\strut{} 1.8}}%
\put(1650,2945){\makebox(0,0)[r]{\strut{} 1.9}}%
\put(1650,3560){\makebox(0,0)[r]{\strut{} 2}}%
\put(1650,4175){\makebox(0,0)[r]{\strut{} 2.1}}%
\put(1650,4790){\makebox(0,0)[r]{\strut{} 2.2}}%
\put(1650,5405){\makebox(0,0)[r]{\strut{} 2.3}}%
\put(1650,6020){\makebox(0,0)[r]{\strut{} 2.4}}%
\put(1650,6635){\makebox(0,0)[r]{\strut{} 2.5}}%
\put(1650,7250){\makebox(0,0)[r]{\strut{} 2.6}}%
\put(1650,7865){\makebox(0,0)[r]{\strut{} 2.7}}%
\put(1650,8480){\makebox(0,0)[r]{\strut{} 2.8}}%
\put(1650,9095){\makebox(0,0)[r]{\strut{} 2.9}}%
\put(1650,9710){\makebox(0,0)[r]{\strut{} 3}}%
\put(1925,550){\makebox(0,0){\strut{} 0}}%
\put(3810,550){\makebox(0,0){\strut{} 10}}%
\put(5695,550){\makebox(0,0){\strut{} 20}}%
\put(7580,550){\makebox(0,0){\strut{} 30}}%
\put(9465,550){\makebox(0,0){\strut{} 40}}%
\put(11350,550){\makebox(0,0){\strut{} 50}}%
\put(13235,550){\makebox(0,0){\strut{} 60}}%
\put(15120,550){\makebox(0,0){\strut{} 70}}%
\put(17005,550){\makebox(0,0){\strut{} 80}}%
\put(18890,550){\makebox(0,0){\strut{} 90}}%
\put(20775,550){\makebox(0,0){\strut{} 100}}%
\put(2114,9403){\makebox(0,0)[l]{\strut{}$T^*$}}%
\put(19267,1407){\makebox(0,0)[l]{\strut{}\large $\left\langle{}m\right\rangle$}}%
\end{picture}%
\endgroup
}\footnotesize\caption{Same as figure \ref{fig5}, but for Shultz
distribution.}\label{fig7}
\end{figure}

Figures \ref{fig5} and \ref{fig7} show shadow curves and critical
binodals in terms of the average chain length
$\left\langle{}m\right\rangle$. Average chain length of the cloud
curve is always 16. In both cases liquid part of the shadow curve
moves to longer chains with the temperature decrease, but for Shultz
distributed system this move is more rapid.

\begin{figure}[h]
\centerline{
\begin{picture}(0,0)%
\includegraphics{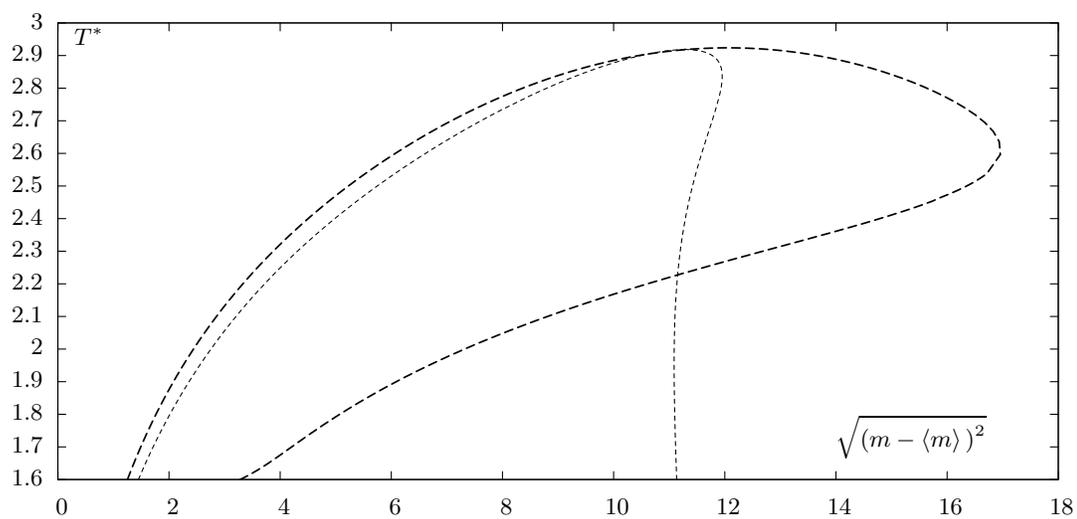}%
\end{picture}%
\begingroup
\setlength{\unitlength}{0.0200bp}%
\begin{picture}(21600,10259)(0,0)%
\put(1650,1100){\makebox(0,0)[r]{\strut{} 1.6}}%
\put(1650,1715){\makebox(0,0)[r]{\strut{} 1.7}}%
\put(1650,2330){\makebox(0,0)[r]{\strut{} 1.8}}%
\put(1650,2945){\makebox(0,0)[r]{\strut{} 1.9}}%
\put(1650,3560){\makebox(0,0)[r]{\strut{} 2}}%
\put(1650,4175){\makebox(0,0)[r]{\strut{} 2.1}}%
\put(1650,4790){\makebox(0,0)[r]{\strut{} 2.2}}%
\put(1650,5405){\makebox(0,0)[r]{\strut{} 2.3}}%
\put(1650,6020){\makebox(0,0)[r]{\strut{} 2.4}}%
\put(1650,6635){\makebox(0,0)[r]{\strut{} 2.5}}%
\put(1650,7250){\makebox(0,0)[r]{\strut{} 2.6}}%
\put(1650,7865){\makebox(0,0)[r]{\strut{} 2.7}}%
\put(1650,8480){\makebox(0,0)[r]{\strut{} 2.8}}%
\put(1650,9095){\makebox(0,0)[r]{\strut{} 2.9}}%
\put(1650,9710){\makebox(0,0)[r]{\strut{} 3}}%
\put(1925,550){\makebox(0,0){\strut{} 0}}%
\put(4019,550){\makebox(0,0){\strut{} 2}}%
\put(6114,550){\makebox(0,0){\strut{} 4}}%
\put(8208,550){\makebox(0,0){\strut{} 6}}%
\put(10303,550){\makebox(0,0){\strut{} 8}}%
\put(12397,550){\makebox(0,0){\strut{} 10}}%
\put(14492,550){\makebox(0,0){\strut{} 12}}%
\put(16586,550){\makebox(0,0){\strut{} 14}}%
\put(18681,550){\makebox(0,0){\strut{} 16}}%
\put(20775,550){\makebox(0,0){\strut{} 18}}%
\put(2239,9403){\makebox(0,0)[l]{\strut{}$T^*$}}%
\put(16586,1899){\makebox(0,0)[l]{\strut{}$\sqrt{\left(m-\left\langle{}m\right\rangle{}\right)^2}$}}%
\end{picture}%
\endgroup
}\footnotesize \caption{The same as \ref{fig5}, but reexpressed in
terms of chain length standard deviation
$\sigma_m=\sqrt{\left(m-\left\langle{}m\right\rangle{}\right)^2}$.
}\label{fig6}
\end{figure}

\begin{figure}[h]
\centerline{
\begin{picture}(0,0)%
\includegraphics{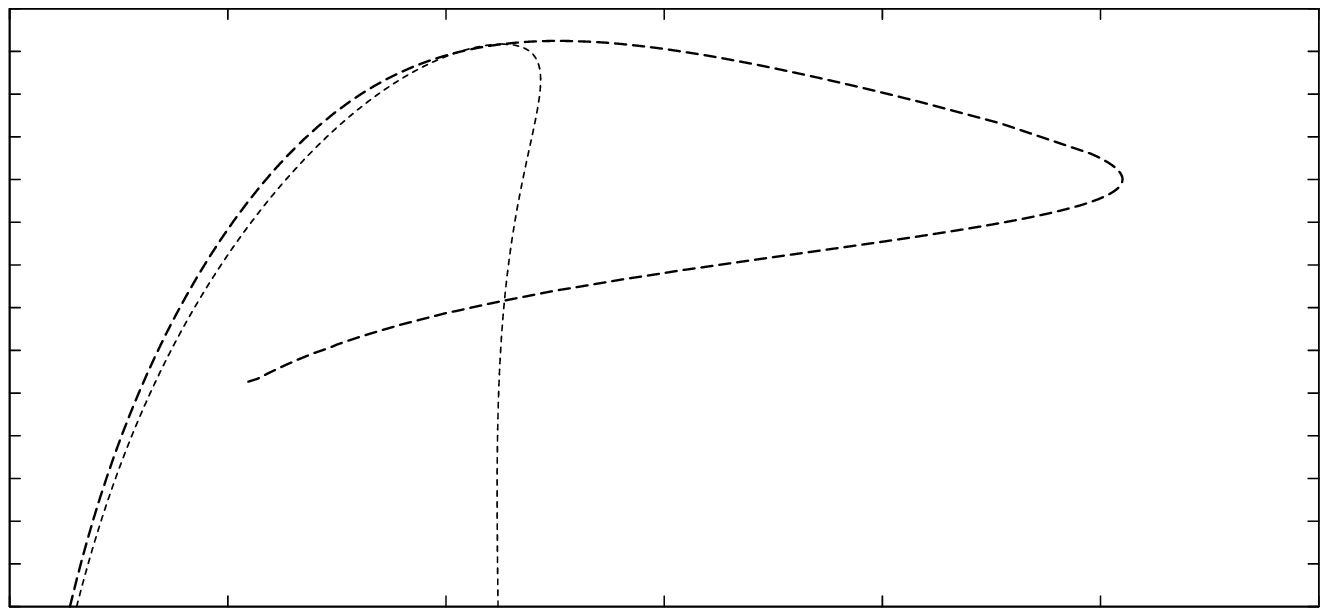}%
\end{picture}%
\begingroup
\setlength{\unitlength}{0.0200bp}%
\begin{picture}(21600,10259)(0,0)%
\put(1650,1100){\makebox(0,0)[r]{\strut{} 1.6}}%
\put(1650,1715){\makebox(0,0)[r]{\strut{} 1.7}}%
\put(1650,2330){\makebox(0,0)[r]{\strut{} 1.8}}%
\put(1650,2945){\makebox(0,0)[r]{\strut{} 1.9}}%
\put(1650,3560){\makebox(0,0)[r]{\strut{} 2}}%
\put(1650,4175){\makebox(0,0)[r]{\strut{} 2.1}}%
\put(1650,4790){\makebox(0,0)[r]{\strut{} 2.2}}%
\put(1650,5405){\makebox(0,0)[r]{\strut{} 2.3}}%
\put(1650,6020){\makebox(0,0)[r]{\strut{} 2.4}}%
\put(1650,6635){\makebox(0,0)[r]{\strut{} 2.5}}%
\put(1650,7250){\makebox(0,0)[r]{\strut{} 2.6}}%
\put(1650,7865){\makebox(0,0)[r]{\strut{} 2.7}}%
\put(1650,8480){\makebox(0,0)[r]{\strut{} 2.8}}%
\put(1650,9095){\makebox(0,0)[r]{\strut{} 2.9}}%
\put(1650,9710){\makebox(0,0)[r]{\strut{} 3}}%
\put(1925,550){\makebox(0,0){\strut{} 0}}%
\put(5067,550){\makebox(0,0){\strut{} 5}}%
\put(8208,550){\makebox(0,0){\strut{} 10}}%
\put(11350,550){\makebox(0,0){\strut{} 15}}%
\put(14492,550){\makebox(0,0){\strut{} 20}}%
\put(17633,550){\makebox(0,0){\strut{} 25}}%
\put(20775,550){\makebox(0,0){\strut{} 30}}%
\put(2302,9403){\makebox(0,0)[l]{\strut{}$T^*$}}%
\put(16377,1899){\makebox(0,0)[l]{\strut{}$\sqrt{\left(m-\left\langle{}m\right\rangle{}\right)^2}$}}%
\end{picture}%
\endgroup
}\footnotesize\caption{Same as figure \ref{fig6}, but for Shultz
distribution.}\label{fig8}
\end{figure}

\begin{figure}[h]
\centerline{
\begin{picture}(0,0)%
\includegraphics{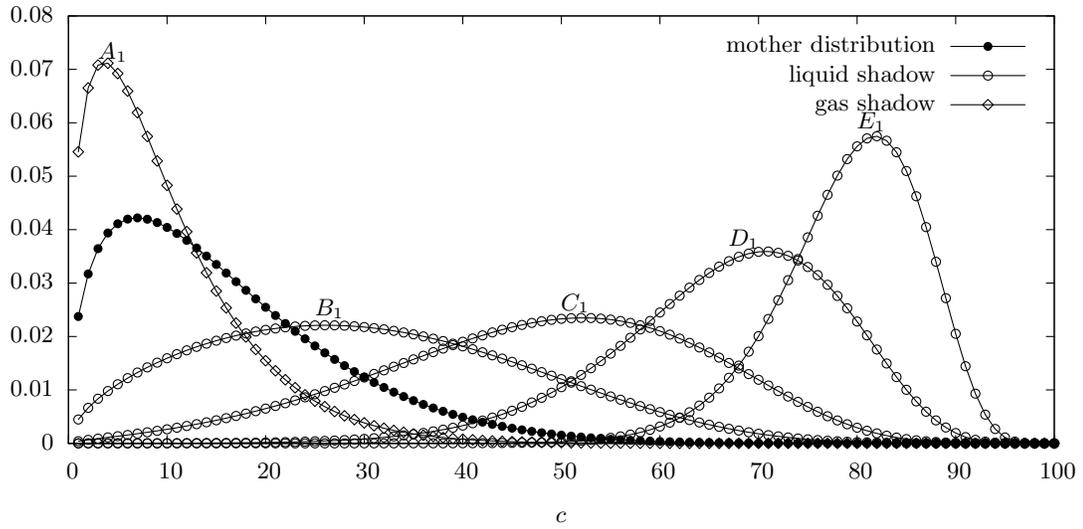}%
\end{picture}%
\begingroup
\setlength{\unitlength}{0.0200bp}%
\begin{picture}(21600,10259)(0,0)%
\put(1925,1650){\makebox(0,0)[r]{\strut{} 0}}%
\put(1925,2658){\makebox(0,0)[r]{\strut{} 0.01}}%
\put(1925,3665){\makebox(0,0)[r]{\strut{} 0.02}}%
\put(1925,4672){\makebox(0,0)[r]{\strut{} 0.03}}%
\put(1925,5680){\makebox(0,0)[r]{\strut{} 0.04}}%
\put(1925,6688){\makebox(0,0)[r]{\strut{} 0.05}}%
\put(1925,7695){\makebox(0,0)[r]{\strut{} 0.06}}%
\put(1925,8703){\makebox(0,0)[r]{\strut{} 0.07}}%
\put(1925,9710){\makebox(0,0)[r]{\strut{} 0.08}}%
\put(2200,1100){\makebox(0,0){\strut{} 0}}%
\put(4058,1100){\makebox(0,0){\strut{} 10}}%
\put(5915,1100){\makebox(0,0){\strut{} 20}}%
\put(7773,1100){\makebox(0,0){\strut{} 30}}%
\put(9630,1100){\makebox(0,0){\strut{} 40}}%
\put(11488,1100){\makebox(0,0){\strut{} 50}}%
\put(13345,1100){\makebox(0,0){\strut{} 60}}%
\put(15203,1100){\makebox(0,0){\strut{} 70}}%
\put(17060,1100){\makebox(0,0){\strut{} 80}}%
\put(18918,1100){\makebox(0,0){\strut{} 90}}%
\put(20775,1100){\makebox(0,0){\strut{} 100}}%
\put(11487,275){\makebox(0,0){\strut{}$c$}}%
\put(2757,9005){\makebox(0,0)[l]{\strut{}$A_1$}}%
\put(6844,4169){\makebox(0,0)[l]{\strut{}$B_1$}}%
\put(11488,4269){\makebox(0,0)[l]{\strut{}$C_1$}}%
\put(14645,5478){\makebox(0,0)[l]{\strut{}$D_1$}}%
\put(17060,7695){\makebox(0,0)[l]{\strut{}$E_1$}}%
\put(18550,9135){\makebox(0,0)[r]{\strut{}mother distribution}}%
\put(18550,8585){\makebox(0,0)[r]{\strut{}liquid shadow}}%
\put(18550,8035){\makebox(0,0)[r]{\strut{}gas shadow}}%
\end{picture}%
\endgroup
}\footnotesize\caption{Distribution functions $f_c=\rho_c/\rho$ of
Beta distributed system of five points, marked on \ref{fig1},
along shadow curve, together with cloud curve (mother)
distribution.}\label{fig9}
\end{figure}

\begin{figure}[h]
\centerline{
\begin{picture}(0,0)%
\includegraphics{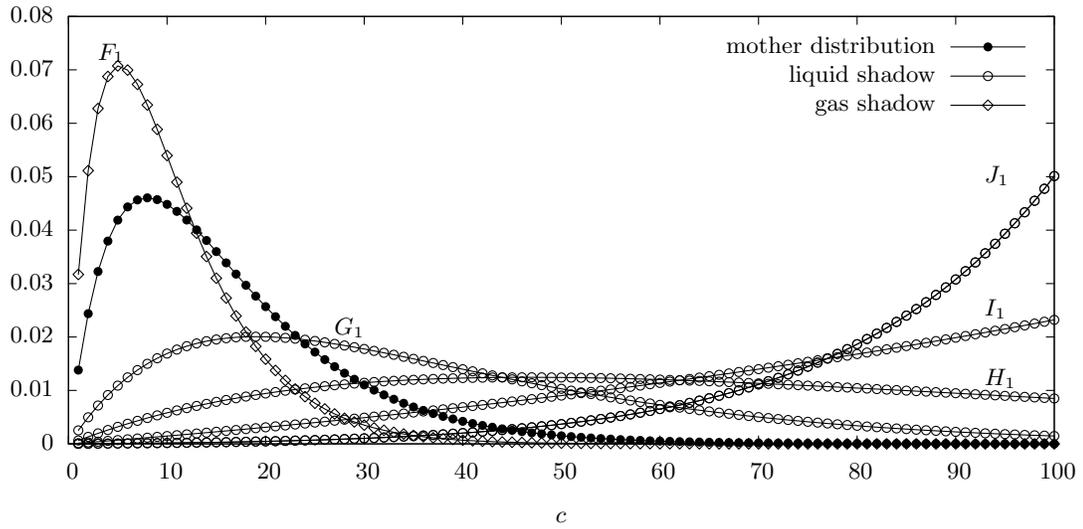}%
\end{picture}%
\begingroup
\setlength{\unitlength}{0.0200bp}%
\begin{picture}(21600,10259)(0,0)%
\put(1925,1650){\makebox(0,0)[r]{\strut{} 0}}%
\put(1925,2658){\makebox(0,0)[r]{\strut{} 0.01}}%
\put(1925,3665){\makebox(0,0)[r]{\strut{} 0.02}}%
\put(1925,4672){\makebox(0,0)[r]{\strut{} 0.03}}%
\put(1925,5680){\makebox(0,0)[r]{\strut{} 0.04}}%
\put(1925,6688){\makebox(0,0)[r]{\strut{} 0.05}}%
\put(1925,7695){\makebox(0,0)[r]{\strut{} 0.06}}%
\put(1925,8703){\makebox(0,0)[r]{\strut{} 0.07}}%
\put(1925,9710){\makebox(0,0)[r]{\strut{} 0.08}}%
\put(2200,1100){\makebox(0,0){\strut{} 0}}%
\put(4058,1100){\makebox(0,0){\strut{} 10}}%
\put(5915,1100){\makebox(0,0){\strut{} 20}}%
\put(7773,1100){\makebox(0,0){\strut{} 30}}%
\put(9630,1100){\makebox(0,0){\strut{} 40}}%
\put(11488,1100){\makebox(0,0){\strut{} 50}}%
\put(13345,1100){\makebox(0,0){\strut{} 60}}%
\put(15203,1100){\makebox(0,0){\strut{} 70}}%
\put(17060,1100){\makebox(0,0){\strut{} 80}}%
\put(18918,1100){\makebox(0,0){\strut{} 90}}%
\put(20775,1100){\makebox(0,0){\strut{} 100}}%
\put(11487,275){\makebox(0,0){\strut{}$c$}}%
\put(2757,9005){\makebox(0,0)[l]{\strut{}$F_1$}}%
\put(7215,3816){\makebox(0,0)[l]{\strut{}$G_1$}}%
\put(19475,2859){\makebox(0,0)[l]{\strut{}$H_1$}}%
\put(19475,4169){\makebox(0,0)[l]{\strut{}$I_1$}}%
\put(19475,6688){\makebox(0,0)[l]{\strut{}$J_1$}}%
\put(18550,9135){\makebox(0,0)[r]{\strut{}mother distribution}}%
\put(18550,8585){\makebox(0,0)[r]{\strut{}liquid shadow}}%
\put(18550,8035){\makebox(0,0)[r]{\strut{}gas shadow}}%
\end{picture}%
\endgroup
}\footnotesize \caption{Same as figure \ref{fig9}, but for system
with Shultz distribution, and phase diagram, containing the
points, given on figure \ref{fig2}} \label{fig10}
\end{figure}

Figures \ref{fig6} and \ref{fig8} expose shadow curves
and critical binodals in terms of the chain length standard deviation
$\sigma_m$. Similarly to the average lengths liquid part of the shadow
curve is on the right and as the temperature decreases
it moves to larger chain length deviations. However after reaching
some maximum it starts to decrease. This decrease corresponds to
the formation and growing of the peak on the distribution function in the region
of long chains. Beta distributed system shows noticeable peak for the
point $E_1$ on the phase diagram, its distribution function can be
found on figure \ref{fig9}. With the temperature decrease the peak
grows and approaches the right margin $M=100$ of chain lengths. The
peak for the shadow phase of Shultz distributed system has irregular
form, its maximum is always at $c=M=100$ (see figure \ref{fig10}
for point $J_1$) and it also grows with the temperature decrease.
These effects indicate that for the low temperatures liquid shadow
phase becoming more and more monodisperse and is composed
mainly of long chains. The same happends for the gas shadow
phase, but it "likes" more short chains than long.

To study conditions at which one can observe irregular peak
formation, we examined six additional phase diagrams of Shultz
distributed system with $z=5$, $z=10$ and at three values of the cutoff
chain lengths, i.e. the number of species $m_{cutoff}=M=100,200,300$.
They can be found at figure 11. We can deduce that increase of
the cutoff chain length lowers temperature of the peak formation. However this
lowering is rather negligible when compared to the change of the peak
appearance temperature due to the change of the mother distribution
polydispersity. Thus, to get the peak on the distribution function of the liquid shadow
phase at high enough temperature one should study the system with high enough polydispersity.

We also examined the change of the critical point position due to
the change of the mother phase distribution function polydispersity (figures \ref{fig12} and \ref{fig13}).
For both systems critical point moves to the higher temperatures and
to the lower volume fractions with the increase of the distribution function width.

\begin{figure}[h]
\centerline{
\begin{picture}(0,0)%
\includegraphics{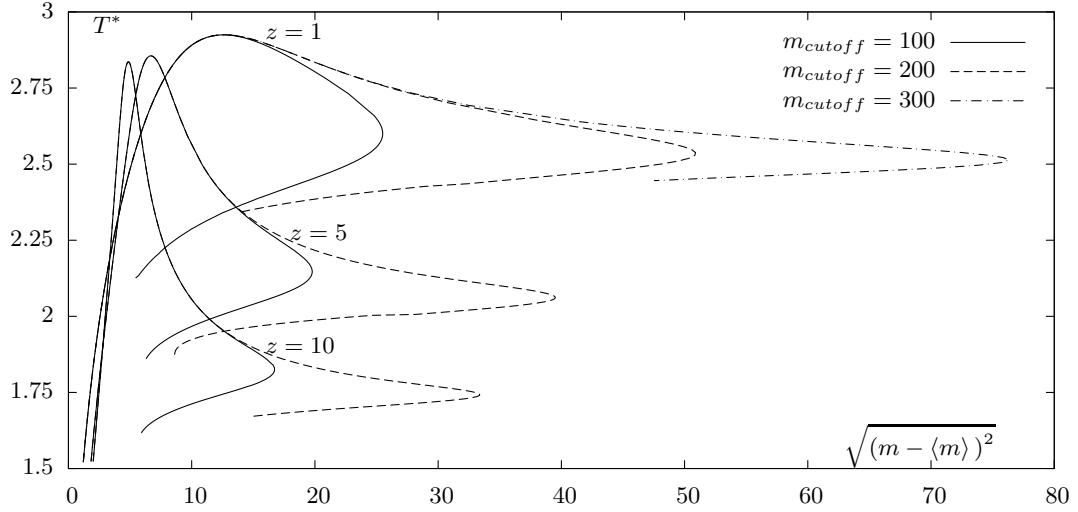}%
\end{picture}%
\begingroup
\setlength{\unitlength}{0.0200bp}%
\begin{picture}(21600,10259)(0,0)%
\put(1925,1100){\makebox(0,0)[r]{\strut{} 1.5}}%
\put(1925,2535){\makebox(0,0)[r]{\strut{} 1.75}}%
\put(1925,3970){\makebox(0,0)[r]{\strut{} 2}}%
\put(1925,5405){\makebox(0,0)[r]{\strut{} 2.25}}%
\put(1925,6840){\makebox(0,0)[r]{\strut{} 2.5}}%
\put(1925,8275){\makebox(0,0)[r]{\strut{} 2.75}}%
\put(1925,9710){\makebox(0,0)[r]{\strut{} 3}}%
\put(2200,550){\makebox(0,0){\strut{} 0}}%
\put(4522,550){\makebox(0,0){\strut{} 10}}%
\put(6844,550){\makebox(0,0){\strut{} 20}}%
\put(9166,550){\makebox(0,0){\strut{} 30}}%
\put(11488,550){\makebox(0,0){\strut{} 40}}%
\put(13809,550){\makebox(0,0){\strut{} 50}}%
\put(16131,550){\makebox(0,0){\strut{} 60}}%
\put(18453,550){\makebox(0,0){\strut{} 70}}%
\put(20775,550){\makebox(0,0){\strut{} 80}}%
\put(5915,9308){\makebox(0,0)[l]{\strut{}$z=1$}}%
\put(6379,5520){\makebox(0,0)[l]{\strut{}$z=5$}}%
\put(5915,3396){\makebox(0,0)[l]{\strut{}$z=10$}}%
\put(2664,9423){\makebox(0,0)[l]{\strut{}$T^*$}}%
\put(16828,1530){\makebox(0,0)[l]{\strut{}$\sqrt{\left(m-\left\langle{}m\right\rangle{}\right)^2}$}}%
\put(18550,9135){\makebox(0,0)[r]{\strut{}$m_{cutoff}=100$}}%
\put(18550,8585){\makebox(0,0)[r]{\strut{}$m_{cutoff}=200$}}%
\put(18550,8035){\makebox(0,0)[r]{\strut{}$m_{cutoff}=300$}}%
\end{picture}%
\endgroup
}\footnotesize \caption{Figure shows shadow curves of system with
Shultz distribution for $z=1$,$z=5$ and $z=10$ and different
cutoffs of distribution function, i.e. different maximal species
chain lengths $100$,$200$ and $300$. The figure is displayed in
terms of $T^*$ and standard deviation of chain length
$\sigma_m=\sqrt{\left(m-\left\langle{}m\right\rangle{}\right)^2}$.
The decrease of chain length deviation with decrease of
temperature that is located below chain length deviation maximum,
corresponds to formation of distribution function peak near the
cutoff distance. The peak appears in liquid branch of shadow
curve. }\label{fig11}
\end{figure}

\begin{figure}[h]
\centerline{
\begin{picture}(0,0)%
\includegraphics{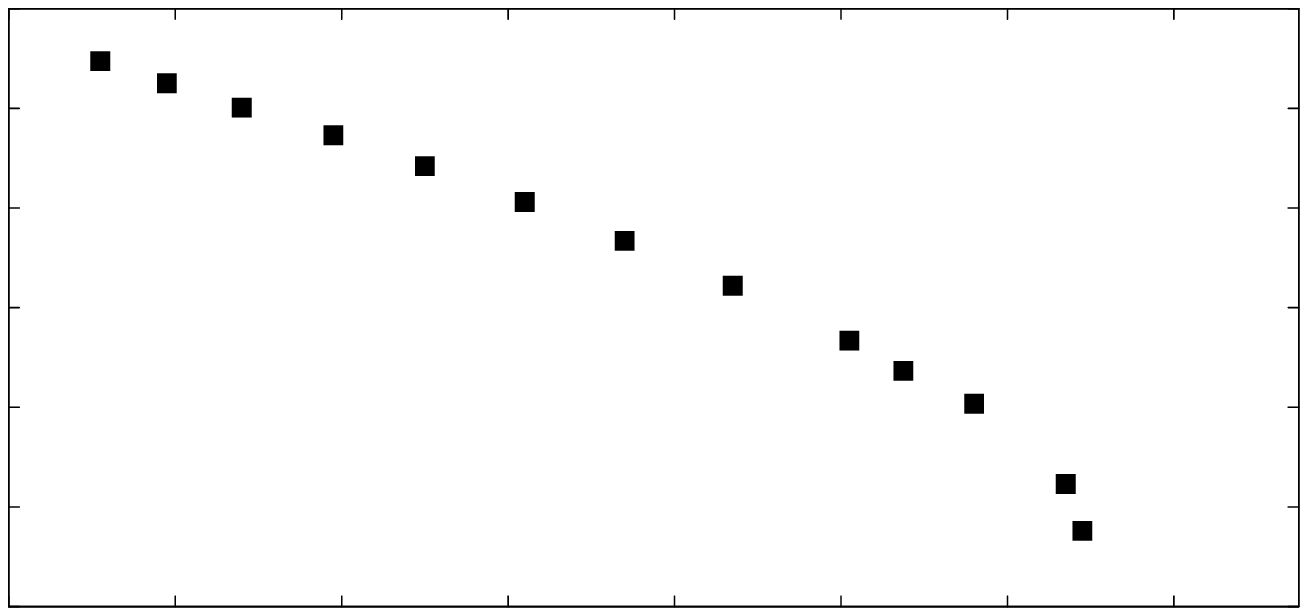}%
\end{picture}%
\begingroup
\setlength{\unitlength}{0.0200bp}%
\begin{picture}(21600,10259)(0,0)%
\put(1925,1100){\makebox(0,0)[r]{\strut{} 2.8}}%
\put(1925,2535){\makebox(0,0)[r]{\strut{} 2.85}}%
\put(1925,3970){\makebox(0,0)[r]{\strut{} 2.9}}%
\put(1925,5405){\makebox(0,0)[r]{\strut{} 2.95}}%
\put(1925,6840){\makebox(0,0)[r]{\strut{} 3}}%
\put(1925,8275){\makebox(0,0)[r]{\strut{} 3.05}}%
\put(1925,9710){\makebox(0,0)[r]{\strut{} 3.1}}%
\put(2200,550){\makebox(0,0){\strut{} 0.086}}%
\put(4597,550){\makebox(0,0){\strut{} 0.088}}%
\put(6994,550){\makebox(0,0){\strut{} 0.09}}%
\put(9390,550){\makebox(0,0){\strut{} 0.092}}%
\put(11787,550){\makebox(0,0){\strut{} 0.094}}%
\put(14184,550){\makebox(0,0){\strut{} 0.096}}%
\put(16581,550){\makebox(0,0){\strut{} 0.098}}%
\put(18977,550){\makebox(0,0){\strut{} 0.1}}%
\put(3818,8955){\makebox(0,0)[l]{\strut{}$\sigma_m$= 23.73}}%
\put(4777,8637){\makebox(0,0)[l]{\strut{}$\sigma_m$= 22.63}}%
\put(5855,8286){\makebox(0,0)[l]{\strut{}$\sigma_m$= 21.47}}%
\put(7173,7888){\makebox(0,0)[l]{\strut{}$\sigma_m$= 20.24}}%
\put(8492,7443){\makebox(0,0)[l]{\strut{}$\sigma_m$= 18.93}}%
\put(9930,6926){\makebox(0,0)[l]{\strut{}$\sigma_m$= 17.53}}%
\put(11368,6366){\makebox(0,0)[l]{\strut{}$\sigma_m$= 16.00}}%
\put(12926,5721){\makebox(0,0)[l]{\strut{}$\sigma_m$= 14.31}}%
\put(14603,4931){\makebox(0,0)[l]{\strut{}$\sigma_m$= 12.39}}%
\put(15382,4495){\makebox(0,0)[l]{\strut{}$\sigma_m$= 11.31}}%
\put(16401,4022){\makebox(0,0)[l]{\strut{}$\sigma_m$= 10.12}}%
\put(17719,2865){\makebox(0,0)[l]{\strut{}$\sigma_m$= 7.16}}%
\put(17959,2191){\makebox(0,0)[l]{\strut{}$\sigma_m$= 5.06}}%
\put(2500,7557){\makebox(0,0)[l]{\strut{}$T^*$}}%
\put(15382,1531){\makebox(0,0)[l]{\strut{}\Large $\eta$}}%
\end{picture}%
\endgroup
}\footnotesize \caption{Critical temperatures and volume fractions
$\eta$ for variety of Beta distributions
dispersions.}\label{fig12}
\end{figure}

\begin{figure}[h]
\centerline{
\begin{picture}(0,0)%
\includegraphics{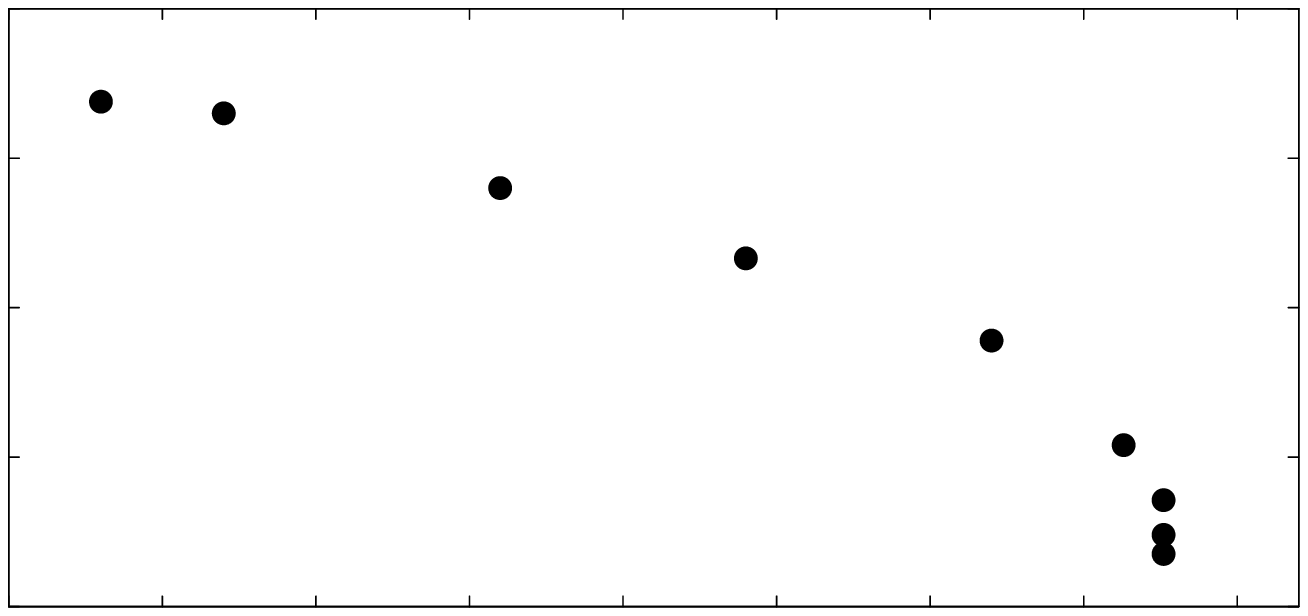}%
\end{picture}%
\begingroup
\setlength{\unitlength}{0.0200bp}%
\begin{picture}(21600,10259)(0,0)%
\put(1925,1100){\makebox(0,0)[r]{\strut{} 2.8}}%
\put(1925,3252){\makebox(0,0)[r]{\strut{} 2.85}}%
\put(1925,5405){\makebox(0,0)[r]{\strut{} 2.9}}%
\put(1925,7557){\makebox(0,0)[r]{\strut{} 2.95}}%
\put(1925,9710){\makebox(0,0)[r]{\strut{} 3}}%
\put(2200,550){\makebox(0,0){\strut{} 0.095}}%
\put(4411,550){\makebox(0,0){\strut{} 0.0955}}%
\put(6623,550){\makebox(0,0){\strut{} 0.096}}%
\put(8834,550){\makebox(0,0){\strut{} 0.0965}}%
\put(11045,550){\makebox(0,0){\strut{} 0.097}}%
\put(13257,550){\makebox(0,0){\strut{} 0.0975}}%
\put(15468,550){\makebox(0,0){\strut{} 0.098}}%
\put(17679,550){\makebox(0,0){\strut{} 0.0985}}%
\put(19890,550){\makebox(0,0){\strut{} 0.099}}%
\put(3836,8634){\makebox(0,0)[l]{\strut{}z=0.05}}%
\put(5605,8203){\makebox(0,0)[l]{\strut{}z=0.1}}%
\put(9586,7127){\makebox(0,0)[l]{\strut{}z=0.5}}%
\put(13124,6115){\makebox(0,0)[l]{\strut{}z=1}}%
\put(16662,4931){\makebox(0,0)[l]{\strut{}z=2}}%
\put(18519,3425){\makebox(0,0)[l]{\strut{}z=5}}%
\put(19139,2633){\makebox(0,0)[l]{\strut{}z=10}}%
\put(19139,2133){\makebox(0,0)[l]{\strut{}z=20}}%
\put(19139,1642){\makebox(0,0)[l]{\strut{}z=40}}%
\put(2642,9280){\makebox(0,0)[l]{\strut{}$T^*$}}%
\put(16574,1444){\makebox(0,0)[l]{\strut{}\Large $\eta$}}%
\end{picture}%
\endgroup
}\footnotesize \caption{Critical temperatures and volume fractions
$\eta$ for variety of Shultz distributions
dispersions.}\label{fig13}
\end{figure}

In our previous work \cite{Kalyuzhnyi2006}, which is dedicated to the Yukawa
hard spheres with size polydispersity we concluded that hard sphere repulsive
interactions plays a major role in controlling fractionation
effects. In the present case the situation is different: strong intramolecular
attraction may change effective radius of the chain molecules,
which are flexible and this may have influence
on the phase behavior. But our theory takes structural properties of
the system only on the level of hard sphere chains without attraction,
and hence it can't capture the effects of effective radius change.

To verify the accuracy of our theory we need to
perform molecular MC simulations similar to those made by Wilding
at al \cite{Nigel2004,Nigel2006}.

\section{Concluding remarks}

We have shown that polydisperse Yukawa chain mixture
with chain length polidispersity treated within HTA belongs to the TFE
class of the models, i.e. its thermodynamical properties were shown to depend
on the finite set of generalized distribution function moments.
Exploiting this property we represent coexistence
relations as a set of four nonlinear equations for unknown moments
of daughter distribution functions. To illustrate how tails of
the distribution functions can impact the phase behavior these equations were solved
and the phase diagrams of the systems with two different
mother phase distribution functions with the same dispersion
characteristics (mean chain length and standard deviation) and
different asymptotics of the distribution function tails were studied. The obtained results show
strong dependence of the shadow curve position on the asymptotic of the mother phase distribution
function.

We also demonstrated that temperature of the appearance of the peak on the
distribution function of the shadow phase strongly depends on the system polydispersity.

\appendix
\section{Expressions for the chemical potential and pressure}

The quantities given below can be simplified, but they are kept her in
the form that resembles the structure of the overall RDFs Laplace
transform.

Contribution from HTA free energy term to chemical potential reads
\bea
\sbspc&\sbspc&\beta\mu_{c,HTA}=-\frac{4\beta\epsilon_0r_0}{\pi{}\Delta D}\left\{ll_c\left[\frac{1}{{z^{(0)}}^2}+\frac{R}{{z^{(0)}}}-\frac{R^3}{\Delta{}{z^{(0)}}}\frac{\varphi_1(R)}{\varphi_0(R)}v+\frac{lR^4}{4\Delta{}{z^{(0)}}}+\frac{R^3}{2\Delta{}{z^{(0)}}^2}\left(l+2v\right)\right]\right.\nonumber\\
\sbspc&\sbspc&\left.+2\left(vl_c+lv_c\right)\left[\frac{1}{{z^{(0)}}^2}+\frac{R}{{z^{(0)}}}-\frac{R}{\varphi_0(R)}\left(\frac{1}{{z^{(0)}}^2}+\frac{R}{2{z^{(0)}}}\right)-\frac{R^3}{\Delta{}{z^{(0)}}}\frac{\varphi_1(R)}{\varphi_0(R)}v+\frac{lR^4}{4\Delta{}{z^{(0)}}}+\frac{\left(l+2v\right)R^3}{2\Delta{}{z^{(0)}}}\frac{\varphi_1(R)}{\varphi_0(R)}\right]\right.\nonumber\\
\sbspc&\sbspc&\left.+4vv_c\left[\left(\frac{\varphi_1(R)}{\varphi_0(R)}\right)^2+R\frac{\varphi_1(R)}{\varphi_0(R)}+\frac{lR^4}{4\Delta{}{z^{(0)}}}+\frac{l+2v}{\Delta}\left(\left(\frac{R^3}{2}+\frac{\varphi_1(R){z^{(0)}}R^2}{2}\right)\left(\frac{\varphi_1(R)}{\varphi_0(R)}\right)^2-R^2\varphi_1(R)\frac{\varphi_1(R)}{\varphi_0(R)}\right)\right.\right.\nonumber\\
\sbspc&\sbspc&\left.\left.+\frac{\left(l+2v\frac{\varphi_1(R)}{\varphi_0(R)}{z^{(0)}}\right)R^2\varphi_1(R)}{2\Delta{}{z^{(0)}}}\right]\right\}
-\frac{4\beta\epsilon_0r_0}{\pi{}} \frac{v_c}{\varphi_0(R)}\\
\sbspc&\sbspc&+\beta\left(f_{HTA}-f^{intra.}_{HTA}\right)\left(\frac{l_cR^3}{6\Delta}-\frac{\frac{\partial{D}}{\partial{\rho_c}}}{D}\right)\nonumber\\
\sbspc&\sbspc&-\frac{2\beta\epsilon_0r_0}{\pi{}\Delta{}D}
\left\{ll\left[-\frac{v_cR^3}{\Delta{}{z^{(0)}}}\frac{\varphi_1(R)}{\varphi_0(R)}-\frac{vR^6l_c}{6\Delta{}^2{z^{(0)}}}\frac{\varphi_1(R)}{\varphi_0(R)}+\frac{l_cR^4}{4\Delta{}{z^{(0)}}}+\frac{lR^7l_c}{24\Delta{}^2{z^{(0)}}}+\frac{R^3\left(l_c+2v_c\right)}{2\Delta{}{z^{(0)}}^2}+\frac{\left(l+2v\right)R^6l_c}{12\Delta{}^2{z^{(0)}}^2}\right]\right.\nonumber\\
\sbspc&\sbspc&\left.+4lv\left[-\frac{R^3v_c}{\Delta{}{z^{(0)}}}\frac{\varphi_1(R)}{\varphi_0(R)}-\frac{vR^6l_c}{6\Delta{}^2{z^{(0)}}}\frac{\varphi_1(R)}{\varphi_0(R)}+\frac{l_cR^4}{4\Delta{}{z^{(0)}}}+\frac{lR^7l_c}{24\Delta{}^2{z^{(0)}}}+\left(\frac{R^3\left(l_c+2v_c\right)}{2\Delta{}{z^{(0)}}}+\frac{\left(l+2v\right)R^6l_c}{12\Delta{}^2{z^{(0)}}}\right)\frac{\varphi_1(R)}{\varphi_0(R)}\right]\right.\nonumber\\
\sbspc&\sbspc&\left.+4vv\left[\frac{l_cR^4}{4\Delta{}{z^{(0)}}}+\frac{lR^7l_c}{24\Delta{}^2{z^{(0)}}}+\left(\frac{l_c+2v_c}{\Delta}+\frac{(l+2v)R^3l_c}{6\Delta^2}\right)\left(\left(\frac{R^3}{2}+\frac{\varphi_1(R){z^{(0)}}R^2}{2}\right)\left(\frac{\varphi_1(R)}{\varphi_0(R)}\right)^2-R^2\varphi_1(R)\frac{\varphi_1(R)}{\varphi_0(R)}\right)\right.\right.\nonumber\\
\sbspc&\sbspc&\left.\left.+\frac{\left(l_c+2v_c\frac{\varphi_1(R)}{\varphi_0(R)}{z^{(0)}}\right)R^2\varphi_1(R)}{2\Delta{}{z^{(0)}}}+\frac{\left(l+2v\frac{\varphi_1(R)}{\varphi_0(R)}{z^{(0)}}\right)R^5\varphi_1(R)l_c}{12\Delta{}^2{z^{(0)}}}\right]\right\}
+\frac{2\beta\epsilon_0r_0}{\pi{}}
\frac{v_c}{\varphi_0(R)}\nonumber,
\eea
where
\be \beta{}f^{intra.}_{HTA}=-\frac{2\beta\epsilon_0r_0}{\pi{}} \frac{2v}{\varphi_0(R)}
\ee
is contribution from intramolecular Yukawa interactions to free
energy (\ref{f_hta3}) and ${\partial{D}}/{\partial{\rho_c}}$ has
the following form
\bea
\frac{\partial{D}}{\partial{\rho_c}}\sbspc&=\sbspc&-\frac{\varphi_2(R)R^3l_c}{3\Delta^2}\left(l+2v\frac{\varphi_1(R)}{\varphi_0(R)}z^{(0)}\right)\left(1+\frac{R^3}{2\Delta}\left(l+2v\right)\right)-\frac{2}{\Delta}\varphi_2(R)\left[\left(l_c+2v_c\frac{\varphi_1(R)}{\varphi_0(R)}s\right)\left(1+\frac{R^3}{2\Delta}\left(l+2v\right)\right)\right.\nonumber\\
\sbspc&\sbspc&+\left.\left(l+2v\frac{\varphi_1(R)}{\varphi_0(R)}s\right)\left(\frac{R^6l_c}{12\Delta^2}\left(l+2v\right)+\frac{R^3}{2\Delta}\left(l_c+2v_c\right)\right)\right]+\frac{\varphi_2(R)R^6l_c}{3\Delta^3}\left(l+2v\right)2v\frac{\varphi_1(R)}{\varphi_0(R)}z^{(0)}\label{dDdrho}\\
\sbspc&\sbspc&+\frac{R^3\varphi_2(R)}{\Delta^2}\left[\left(l_c+2v_c\right)2v+\left(l+2v\right)2v_c\right]\frac{\varphi_1(R)}{\varphi_0(R)}z^{(0)}
-\frac{R^3l_c}{3\Delta^2}\varphi_1(R)R
\left[\left(l+2v\right)\left(1+\frac{lR^3}{2\Delta}\right)+\left(l+2v\frac{\varphi_1(R)}{\varphi_0(R)}z^{(0)}\right)\right]\nonumber\\
\sbspc&\sbspc&-\frac{1}{\Delta}\varphi_1(R)R
\left[\left(l_c+2v_c\right)\left(1+\frac{lR^3}{2\Delta}\right)+\left(l+2v\right)\left(\frac{R^3l_c}{2\Delta}+\frac{lR^6l_c}{12\Delta^2}\right)+\left(l_c+2v_c\frac{\varphi_1(R)}{\varphi_0(R)}z^{(0)}\right)\right].\nonumber
\eea
Corresponding pressure expression is given by
\bea
\sbspc&\sbspc&\beta{}P_{HTA}=\beta{}f_{HTA}+\beta\left(f_{HTA}-f^{intra.}_{HTA}\right)\left(\frac{lR^3}{6\Delta}-\frac{\sum_c\rho_c\frac{\partial{D}}{\partial{\rho_c}}}{D}\right)\nonumber\\
\sbspc&\sbspc&-\frac{2\beta\epsilon_0r_0}{\pi{}\Delta{}D}
\left\{ll\left[-\frac{vR^3}{\Delta{}{z^{(0)}}}\frac{\varphi_1(R)}{\varphi_0(R)}-\frac{vR^6l}{6\Delta{}^2{z^{(0)}}}\frac{\varphi_1(R)}{\varphi_0(R)}+\frac{lR^4}{4\Delta{}{z^{(0)}}}+\frac{lR^7l}{24\Delta{}^2{z^{(0)}}}+\frac{R^3\left(l+2v\right)}{2\Delta{}{z^{(0)}}^2}+\frac{\left(l+2v\right)R^6l}{12\Delta{}^2{z^{(0)}}^2}\right]\right.\\
\sbspc&\sbspc&\left.+4lv\left[-\frac{R^3v}{\Delta{}{z^{(0)}}}\frac{\varphi_1(R)}{\varphi_0(R)}-\frac{vR^6l}{6\Delta{}^2{z^{(0)}}}\frac{\varphi_1(R)}{\varphi_0(R)}+\frac{lR^4}{4\Delta{}{z^{(0)}}}+\frac{lR^7l}{24\Delta{}^2{z^{(0)}}}+\left(\frac{R^3\left(l+2v\right)}{2\Delta{}{z^{(0)}}}+\frac{\left(l+2v\right)R^6l}{12\Delta{}^2{z^{(0)}}}\right)\frac{\varphi_1(R)}{\varphi_0(R)}\right]\right.\nonumber\\
\sbspc&\sbspc&\left.+4vv\left[\frac{lR^4}{4\Delta{}{z^{(0)}}}+\frac{lR^7l}{24\Delta{}^2{z^{(0)}}}+\left(\frac{l+2v}{\Delta}+\frac{(l+2v)R^3l}{6\Delta^2}\right)\left(\left(\frac{R^3}{2}+\frac{\varphi_1(R){z^{(0)}}R^2}{2}\right)\left(\frac{\varphi_1(R)}{\varphi_0(R)}\right)^2-R^2\varphi_1(R)\frac{\varphi_1(R)}{\varphi_0(R)}\right)\right.\right.\nonumber\\
\sbspc&\sbspc&\left.\left.+\frac{\left(l+2v\frac{\varphi_1(R)}{\varphi_0(R)}{z^{(0)}}\right)R^2\varphi_1(R)}{2\Delta{}{z^{(0)}}}+\frac{\left(l+2v\frac{\varphi_1(R)}{\varphi_0(R)}{z^{(0)}}\right)R^5\varphi_1(R)l}{12\Delta{}^2{z^{(0)}}}\right]\right\}
+\frac{2\beta\epsilon_0r_0}{\pi{}}
\frac{v}{\varphi_0(R)}\nonumber,
\eea
where $\sum_c\rho_c\frac{\partial{D}}{\partial{\rho_c}}$ can be
obtained from (\ref{dDdrho}) by simple replacing of all $l_c$ and
$v_c$ by $l=\sum_c\rho_{c}l_c$ and $v=\sum_c\rho_{c}v_c$.


\begin{references}

\bibitem{my1} Yu.V.Kalyuzhnyi, and G.Kahl, {\it J.Chem.Phys.} {\bf 119}, 7335(2003).
\bibitem{my2} Yu.V.Kalyuzhnyi, G.Kahl, and P.T.Cummings, {\it J.Chem.Phys.}
       {\bf 120}, 10133(2004).
\bibitem{my3} Yu.V.Kalyuzhnyi, G.Kahl, and P.T.Cummings, {\it Europhys.Lett.},
       {\bf 72}, 96(2005).
\bibitem{my4} Yu.V.Kalyuzhnyi, G.Kahl, and P.T.Cummings, {\it J.Chem.Phys.}
       {\bf 123}, 124501(2005).
\bibitem{my5} Yu.V.Kalyuzhnyi, and P.T.Cummings, {\it J.Chem.Phys.}
        {\bf 124}, 114509(2006).
\bibitem{Kalyuzhnyi2006} Yu.V.Kalyuzhnyi, and S.P.Hlushak,
 {\it J.Chem.Phys.} {\bf 125}, 034501(2006).
\bibitem{henderson} J. A. Barker, and D. Henderson, {\it Rev. Mod. Phys.} {\bf 48}, 587(1976).
\bibitem{hansen} J.-P. Hansen, and I. R. McDonald, {\it Theory of Simple Liquids}, Acad. Press, London (1990).
\bibitem{proza1} Yu.V.Kalyuzhnyi, C.-T.Lin, and G.Stell, {\it J.Chem.Phys.} {\bf 106}, 1940(1997);
    ibid {\bf 108}, 6513, 6525(1998),
\bibitem{proza2} G.Stell, C.-T.Lin, and Yu.V.Kalyuzhnyi, {\it J.Chem.Phys.} {\bf 110}, 5444(1999).
\bibitem{proza3} C.-T.Lin, G.Stell, and Yu.V.Kalyuzhnyi, {\it J.Chem.Phys.} {\bf 112}, 3071(2000).
\bibitem{w_jcp} M.S.Wertheim, {\it J.Chem.Phys.} {\bf 87}, 7323(1987).
\bibitem{Alejandro1997} P.J.Whitehead, S.J.Mills, A.Gil-Villegas, A.Galindo, and G.Jackson,
        {\it J.Chem.Phys.} {\bf 106}, 4168(1997).
\bibitem{chang1} J.Chang, and S.I.Sandler, {\it J.Chem.Phys.} {\bf 102}, 437(1995).
\bibitem{chang1} J.Chang, and S.I.Sandler, {\it J.Chem.Phys.} {\bf 102}, 437(1995).
\bibitem{chang2} J.Chang, and S.I.Sandler, {\it J.Chem.Phys.} {\bf 103}, 3196(1995).
\bibitem{Baxter1970} R.J.Baxter, {\it J.Chem.Phys.} {\bf 52}, 4559(1970).
\bibitem{Blum} L.Blum, and J.S.Hoye, {\it J.Chem.Phys.} {\bf 81}, 1311(1977).
\bibitem{Golovko} M.F.Golovko, and I.A.Protsykevich, {\it Chem.Phys.Lett.} {\bf 142}, 463(1987).
\bibitem{Protsykevich} M.F.Holovko, and I.A.Protsykevich, {\it Cond.Matt.Phys.} {\bf 10}, 137(1997).
\bibitem{Kalyuzhnyi2004} E.Whitebay, P.T.Cummings, and Yu.V.Kalyuzhnyi, C.McCabe,
    {\it J.Chem.Phys.} {\bf 121}, 8128(2004).
\bibitem{Mansoori1971} K.E.Starling, G.A.Mansoori, N.F.Carnahan, and Jr.T.W.Leland, {\it J.Chem.Phys.}
    {\bf 54}, 1523(1971).
\bibitem{Bellier-Castella2000} L.Bellier-Castella, H.Xu, and M.Baus, {\bf J.Chem.Phys.} {\bf 113}
     8337(2000).
\bibitem{Nigel2004} M.Fasolo, N.B.Wilding, and P.Sollich, {\it J.Chem.Phys.} {\bf 121}, 6887(2004).
\bibitem{Nigel2006} P.Sollich, N.B.Wilding, M.Buzzacchi, and M.Fasolo, {\it J.Chem.Phys.} {\bf 125},
     014908(2006).


\end{references}

\end{document}